\documentclass[a4paper,10pt]{article}
\newcommand{\pdiff}[2]{\ensuremath{\frac{\partial #1}{\partial #2}}} 
\newcommand{\phia}{\ensuremath{\phi_\alpha}}
\usepackage[english]{babel}
\usepackage[utf8x]{inputenc}
\usepackage{hyperref}
\usepackage{graphicx}
\usepackage{caption}
\usepackage[labelformat=simple]{subcaption}

\usepackage{mathtools}
\usepackage{amsfonts}
\usepackage{xfrac}
\usepackage{siunitx}
\usepackage{placeins}
\usepackage{tikz}
\usepackage{pgfplots}
\usetikzlibrary{calc}
\usetikzlibrary{backgrounds}
\usepgflibrary{shapes}
\usetikzlibrary{through}
\usetikzlibrary{intersections}
\usetikzlibrary{matrix}
\usepackage{multirow}
\usepackage{hyperref}
\usepackage{cleveref}
\crefname{figure}{Fig.}{Figs.}
\Crefname{figure}{Figure}{Figures}

\crefname{table}{Table}{Tables}
\Crefname{table}{Table}{Tables}

\usepackage{xifthen}

\usepackage{esdiff}
\usepackage{ifthen}

\usepackage{booktabs}
\usepackage{nicefrac}

\usepackage[
]{todonotes}
\presetkeys{todonotes}{inline}{}
\usepackage{footmisc}

\usepackage[singlespacing]{setspace}

\newcommand{\rz}{{\if mm {\rm I}\mkern -3mu{\rm R}\else \leavevmode
\hbox{I}\kern -.17em \hbox{R} \fi}}
\newcommand{\nz}{{\if mm {\rm I}\mkern -3mu{\rm N}\else \leavevmode
\hbox{I}\kern -.17em \hbox{N} \fi}}

\newcommand{\Vap}{\ensuremath{\mathrm{V}}}

\newcommand{\grad}[1]{\ensuremath{\nabla{#1}}}
\renewcommand{\div}[1]{\ensuremath{\nabla\cdot{#1}}}

\newcommand{\ha}{\ensuremath{h_\alpha}}

\newcommand{\contact}{\ensuremath{\mathrm{C}}}

\newdimen\CdotAxis
\newcommand*{\CdotAux}[3]{%
  {%
    \settoheight\CdotAxis{$#2\vcenter{}$}%
    \sbox0{%
      \raisebox\CdotAxis{%
        \scalebox{#1}{%
          \raisebox{-\CdotAxis}{%
            $\mathsurround=0pt #2#3$%
          }%
        }%
      }%
    }%
    \dp0=0pt %
    \sbox2{$#2\bullet$}%
    \ifdim\ht2<\ht0 %
      \ht0=\ht2 %
    \fi
    \sbox2{$\mathsurround=0pt #2#3$}%
    \hbox to \wd2{\hss\usebox{0}\hss}%
  }%
}

\DeclareSIUnit{\molpc}{mol\text{-}\%}

\newcommand{\cor}[1]{\textcolor{black}{#1}}

\newcommand{\corM}[1]{\textcolor{black}{#1}}

\providecommand{\keywords}[1]
{
  \small	
  \textbf{\textit{Keywords:}} #1
}

\title{Unravelling densification during sintering by multiscale modelling of grain motion}
\usepackage{authblk}
\author[1,2*]{{Marco Seiz}}

\author[2]{{Henrik Hierl}}
\author[1,2,3]{{Britta Nestler}}

\affil[1]{{Institute of Applied Materials, Karlsruhe Institute of Technology, Stra\ss{}e am Forum 7, 76131 Karlsruhe, Germany}}
\affil[2]{{Institute of Nanotechnology, Karlsruhe Institute of Technology, Hermann-von-Helmholtz-Platz 1, 76344 Eggenstein-Leopoldshafen, Germany}}
\affil[3]{{Institute of Digital Materials,  Karlsruhe University of Applied Sciences, Moltkestr. 30, 76133 Karlsruhe, Germany}}
\affil[*]{{corresponding author: marco.seiz@kit.edu}}

\usepackage[square,numbers,sort&compress]{natbib}

\begin{document}

\maketitle


\begin{abstract}
The resulting microstructure after the sintering process determines many materials properties of interest.
In order to understand the microstructural evolution, simulations are often employed.
One such simulation method is the phase-field method, which has garnered much interest in recent decades.
However, the method lacks a complete model for sintering, as previous works could show unphysical effects and the inability to reach representative volume elements.
Thus the present paper aims to close this gap by employing molecular dynamics and determining rules of motion which can be translated to a phase-field model.
\corM{The key realization is that vacancy absorption induced motion of grains travels through a grain structure without resistance.
Hence the total displacement field of a green body is simply the superposition of all grains reacting in isolation to local vacancy absorption events.}
The resulting phase-field model is shown to be representative starting from particle counts between 97 and 262 and contains the qualitative correct dependence of sintering rate on particle size.
\end{abstract}

\section{Introduction}

The sintering process is an important step in many materials processing routes, from the humble coffee cup to complex applications such as solide oxide fuel cells.
Especially for the latter, the properties of the material are critical for the application.
The materials properties depend on the microstructure of the material and thus predicting the microstructure is of large importance.
While analytical models provide a first avenue to microstructural predictions, the necessary simplifications in geometry and other aspects often prevent quantitative predictions.
Simulations offer another avenue which is less restrictive in the necessary simplifications.
In recent decades, field-resolved simulation methods such as the Monte Carlo method\cite{Cardona2011,Zhang2019} and the phase-field method\cite{Wang2006,Biswas2016,Hoetzer2019,Greenquist2020,Seiz2023a} have attracted much attention for the simulation of sintering.
It was recently shown\cite{Seiz2023a} that the most popular phase-field model of sintering showed a significant influence of the green body size on the densification behaviour.
Since this runs counter to physical intuition and experimental evidence, the present paper aims to eliminate this problem.

The goal of this paper is twofold:
First, qualitative rules of motion during sintering are determined, as to allow the prediction of grain motion and length changes.
This will be achieved by conducting molecular dynamics simulations in geometries which allow the isolation of the relevant processes.
Second, these rules are translated into a phase-field model in order to allow for simulations of arbitrary geometry and scale.
Following this translation, the phase-field model is compared against other phase-field models of sintering before being used to determine representative volume elements.

\section{Molecular dynamics}
Molecular dynamics (MD) is a method in which the dynamics of individual atoms under the influence of an interaction potential can be simulated.
The individual atoms are assumed to follow Newton's laws of motion, with the interaction potential determining what kind of material is being simulated.
Sintering has previously been investigated with MD by various authors\cite{Hawa2007,Hawa2007a,Cheng2013,Ding2009}, but with a general focus on identifying the sintering mechanisms at the nano-scale rather than determining rules of motion for coarser spatial methods.
The present study is somewhat similar to Hawa and Zachariah's work \cite{Hawa2007,Hawa2007a}, in which they investigated how a chain of amorphous Si sintered and considered the influence of chain length and particle size and how these affect the velocity distribution and sintering time.

For the present investigation LAMMPS\cite{Thompson2022} is used to conduct MD simulations.
As a model material copper is employed by using EAM potential developed by Foiles et al. \cite{Foiles1986}.
A comparison with more recent copper EAM potentials was conducted.
While the quantitative results did change, the qualitative trends did not and the employed potential was much faster to calculate.
The timestep employed within the MD simulations is generally \SI{0.004}{ps}.

\begin{figure}
 \includegraphics[width=\columnwidth]{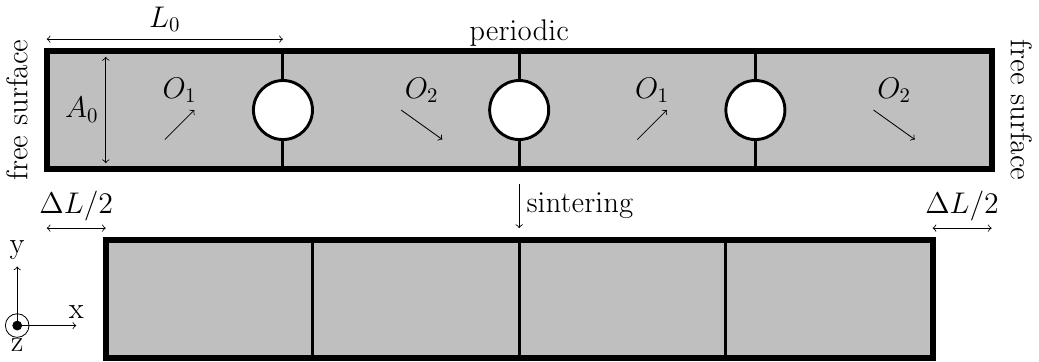}
 \caption{Two-dimensional sketch of the considered geometry in the MD simulations.
 Grains of different orientation $O_1, O_2$ are placed in a row, with pores on grain boundaries.
 The ends of the chain are free surfaces, with directions perpendicular to these being periodic.}
 \label{fig:md-sketch}
\end{figure}

The primary goal of the following simulations is to predict the sample length change $\Delta L$ during sintering, parametrized by variables accessible on the mesoscopic phase-field scale.
For this purpose the geometry depicted in \cref{fig:md-sketch} is developed.
It contains a chain of rectangular cuboid grains of alternating, different orientations, with both ends of the chain being free as to allow movement.
On each of the grain boundaries a pore may be placed, which then vanishes during the sintering process, which in turn induces movement of the grains.
By employing cuboid grains extending to the periodic boundary, the grain rotation as commonly found in MD simulations of sintering\cite{Ding2009} is largely suppressed.
This makes the tracking of the pore region much simpler.
It also makes the calculation of rigid displacements simpler because no rotational displacement needs to be removed.

The following results will be mainly based on placing spherical pores of radius $\approx\SI{1.2}{nm}$, but the qualitative trends of the results do not change when the size is varied or when a cylindrical pore is placed.
The visualization of the results is done with OVITO\cite{ovito} and matplotlib\cite{matplotlib}.
The view of the simulations will be from the positive z direction as indicated in \cref{fig:md-sketch}, unless mentioned otherwise.

The system is prepared as follows:
First, a bicrystal of size $(nL_0$, $mA_0)$ is prepared with a base length $L_0 = \SI{141}{\angstrom}$ and base area $A_0 = \SI{1306}{\angstrom^2}$ with free surfaces along the [100] axis (x-axis) of the simulation cell.
$n$, $m$ are positive integers, which will be varied.
The main grain boundary orientation relationship which will be investigated in the present study is the (210)/[001] STGB.
It is a symmetrical tilt grain boundary (STGB) with a misorientation angle of $\theta = \SI{53.1}{\degree}$ about the [001] tilt axis, with the grain boundary plane being (210).
Additional GBs for which the simulations were conducted are the (310)/[001] STGB and an asymmetrical tilt grain boundary with the left/right grain boundary planes being the (100) and (2$\bar{1}$0) planes, with a \SI{26.56}{\degree} rotation around the [001] axis.
These GBs are chosen because they were easy to directly construct within LAMMPS.
After the atoms are set, a conjugated gradient minimization at $\SI{0}{K}$ is conducted, followed by an constant number of atoms $N$, pressure $p=0$ and temperature $T$ (NPT) ensemble heating run from $\SI{1}{K}$ to the target temperature $T = \SI{700}{K}$ over $\SI{200}{ps}$, followed by another $\SI{320}{ps}$ of equilibration at constant target temperature $T=\SI{700}{K}$.
This condition of $p=0$ and $T=\SI{700}{K}$ will also be held for the rest of the paper.
The system is then copied and shifted until the desired chain length is reached and equilibrated for a period of $\SI{400}{ps}$ to allow the system to relax the newly constructed grain boundaries.
Once the system is ready, regions on the grain boundaries are defined and the atoms removed in order to place pores.
By counting the number of atoms within these regions during the simulation, it is possible to obtain an estimate of the number of absorbed vacancies.
Furthermore, the center of mass (COM) of the individual grains is tracked as to allow calculation of grain displacements and the total length change.
The grain displacements are calculated directly by subtracting the center of mass $x_i(t)$ of grain $i$ at time $t$ from that at time $t=0$.
The total length change in a direction is assumed to be the length change of the vector connecting the center of mass of the first and last grain in the chain.
It is taken to be \emph{positive} for a shortening of the vector.
The sintering simulations on structures with pores are generally run in increments of $\SI{8}{ns}$.
If a pore is observed to have vanished, the simulation run may be terminated early.
Not all simulations are run to complete pore elimination as the goal of the study is to find rules which are also applicable during the process and not only after it.

\begin{figure}
\begin{subfigure}{\textwidth}
\includegraphics[width=\textwidth]{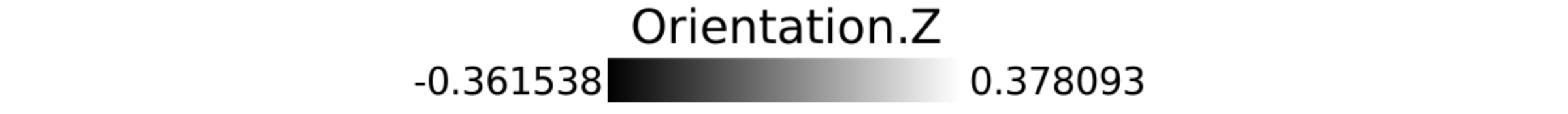} 
\end{subfigure}
\begin{subfigure}[b]{0.3\textwidth}
\includegraphics[width=\textwidth]{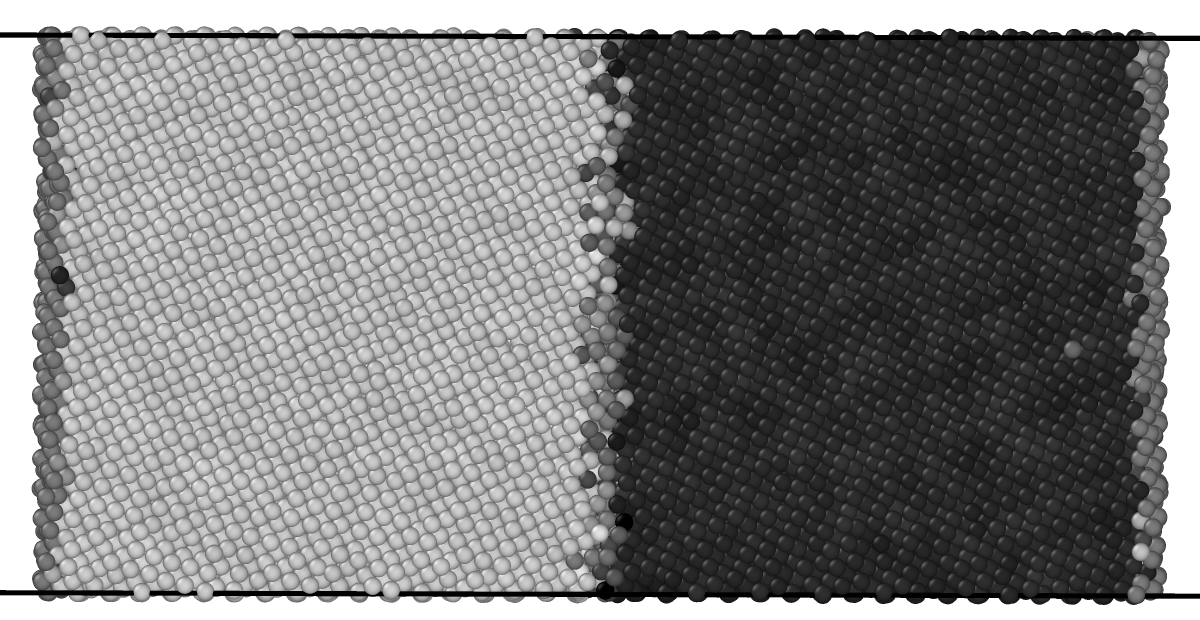}
\caption{initial state showing all atoms}
\label{fig:inip}
\end{subfigure}
~
\begin{subfigure}[b]{0.3\textwidth}
\includegraphics[width=\textwidth]{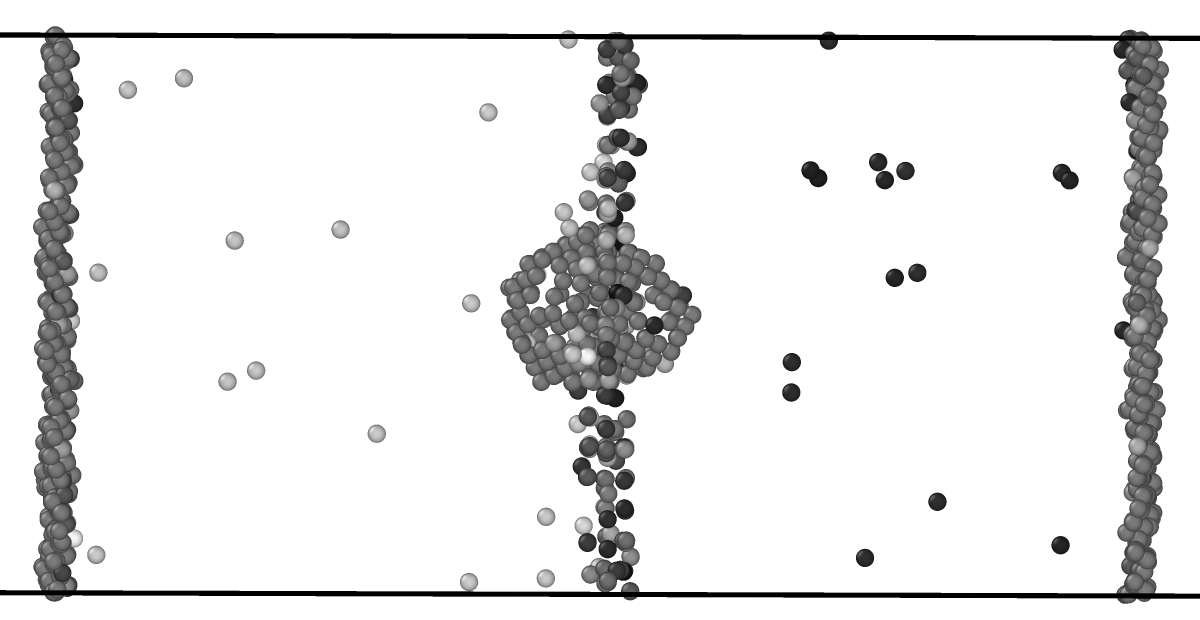}
\caption{intermediate state showing HE atoms}
\label{fig:midp}
\end{subfigure}
~
\begin{subfigure}[b]{0.3\textwidth}
\includegraphics[width=\textwidth]{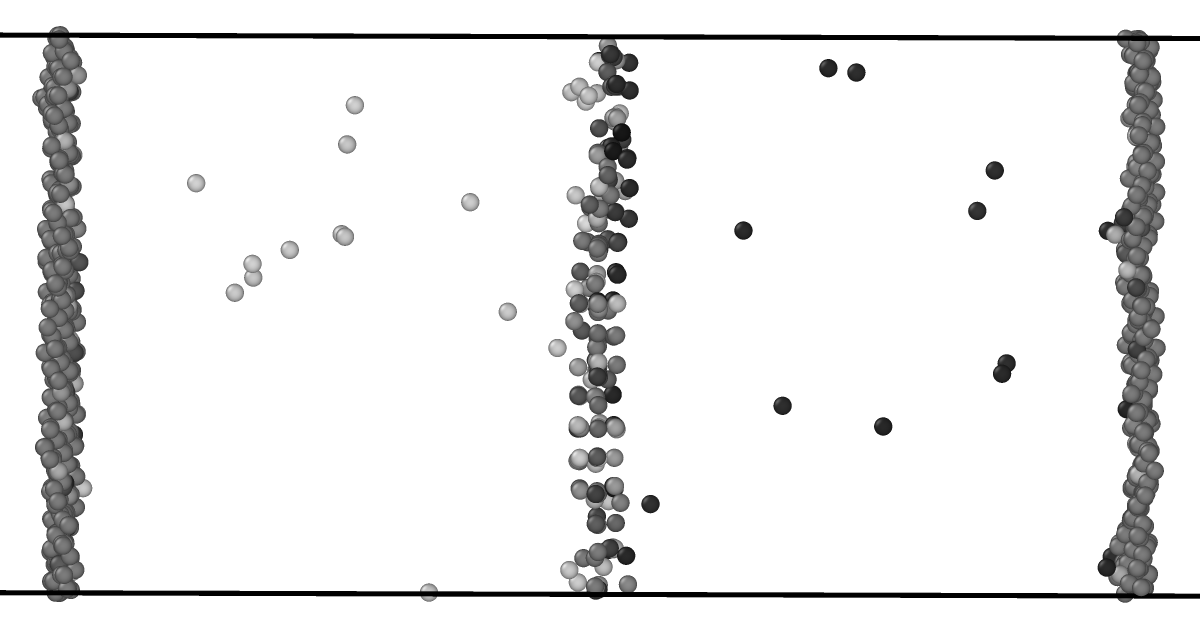} 
\caption{final state showing HE atoms}
\label{fig:endp}
\end{subfigure}

\centering
\begin{subfigure}[]{0.3\textwidth}
\includegraphics[width=\textwidth]{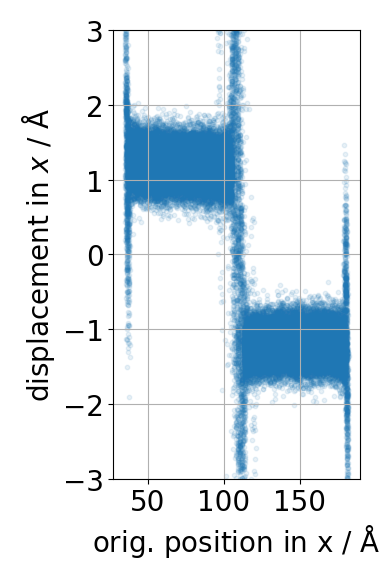} 
\caption{displacement in x}
\end{subfigure}
\begin{subfigure}[]{0.3\textwidth}
\includegraphics[width=\textwidth]{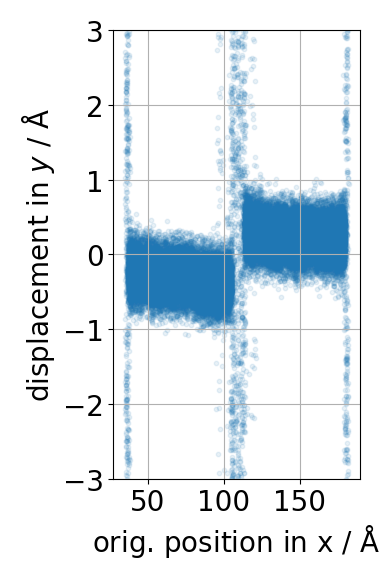} 
\caption{displacement in y}
\end{subfigure}
\begin{subfigure}[]{0.3\textwidth}
\includegraphics[width=\textwidth]{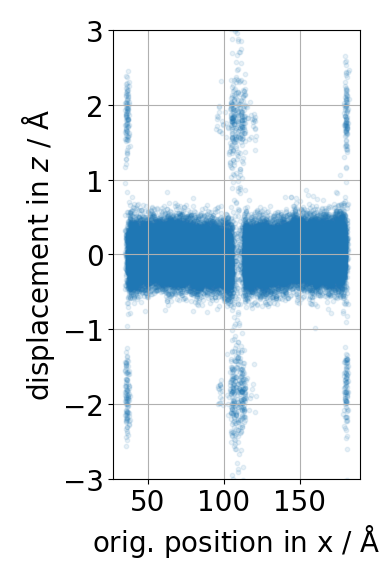} 
\caption{displacement in z}
\end{subfigure}

\caption{
Simulation results exemplarily depicted for $L=2L_0$, $A=2A_0$ and a cylindrical pore.
The color in the top row indicates the local orientation in the viewing direction (Z), calculated via polyhedral template matching\cite{Larsen2016}, allowing to distinguish the grains.
In the last two two images, only high energy (HE) atoms (potential energy of $>\SI{-3.2}{eV}$) are shown, revealing the interfaces.
The displacement of the atoms in the x (d), y (e) and z (f) directions after the pore has vanished are plotted over the original atomic position on the bottom.
The displacement range is fixed to [-3, 3]\si{\angstrom} since the large displacements on the surfaces would obscure the behaviour within the grain.
The inner parts of the grains are generally homogeneously displaced, with the interfaces showing large deviations.
Similar plots are obtained for all other simulations.
}
\label{fig:md-evolution}
\end{figure}

A typical simulation result, starting with a cylindrical pore, is shown in \cref{fig:md-evolution} with the displacement per atom on the bottom.
As expected, the pore vanishes over time and its vanishing is correlated to an inwards movement of the free surfaces.
The displacement per atom is observed to be largely homogeneous within the grains, with large deviations only found close to interfaces, lending credence to the common assumption of rigid-body motion during sintering.
The displacement in x, the shrinkage direction, generally directly scales with the pore size and is influenced by the area of the grains.
While there were differences for the displacement in y, they generally did not show any monotonic relationship to pore size or grain geometry.
In the shown simulation a gradient for the y displacement seems to exist, though this did not always manifest in the other simulations.
Interestingly, the displacement in z did not show a sign difference between the grains.
This might be due to the tilt axis being the z axis, though further research in this direction should be conducted.
For the present paper we shall focus on the displacement in the shrinkage direction and assume that it is described by a rigid-body motion.

\begin{figure}
\begin{center}
\begin{subfigure}[b]{1\textwidth}
\includegraphics[width=\textwidth]{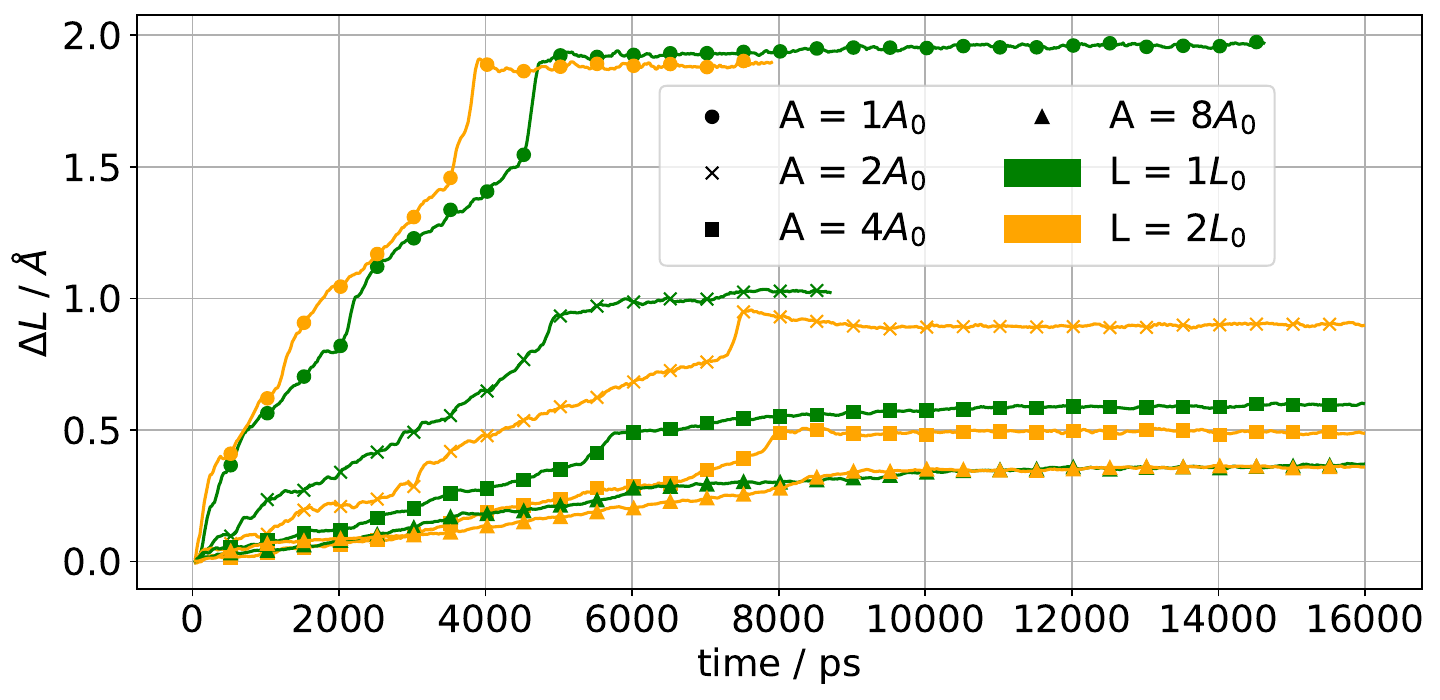}
\end{subfigure}
\begin{subfigure}[b]{1\textwidth}
\includegraphics[width=\textwidth]{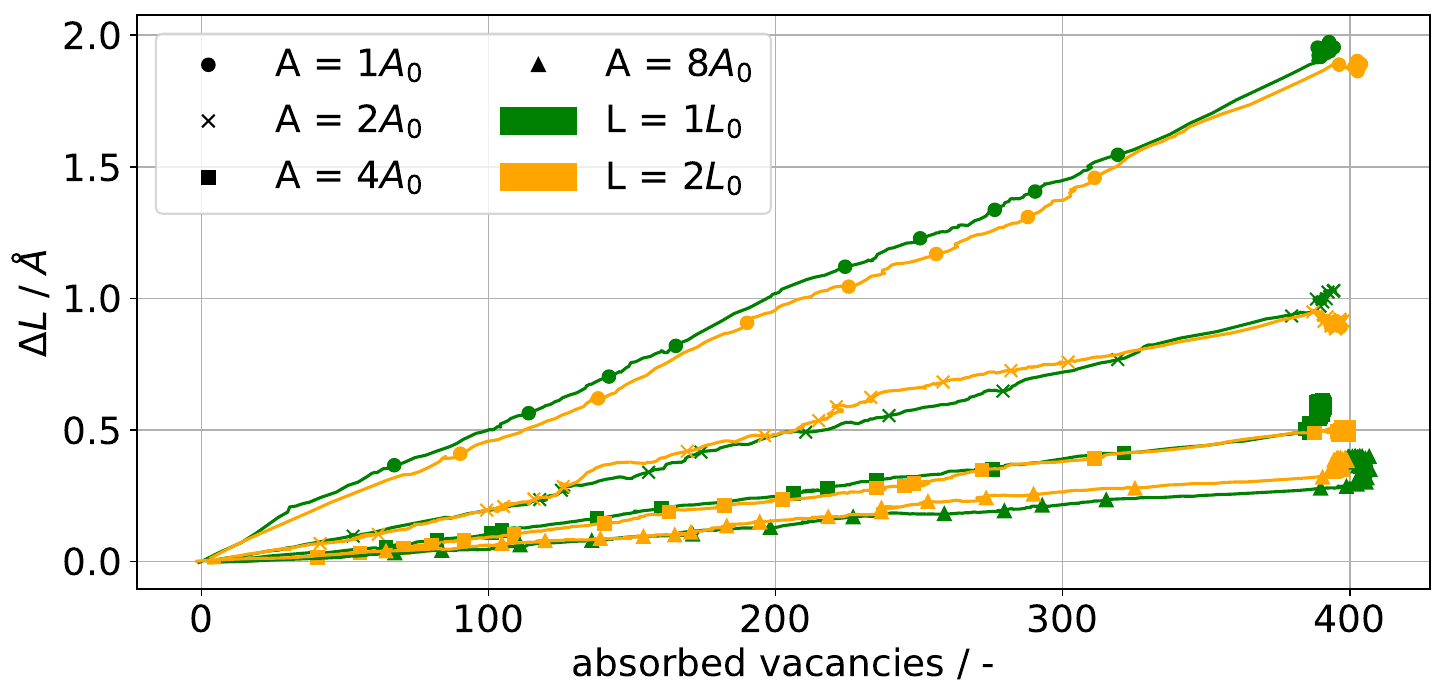} 
\end{subfigure}
\caption{Length change $\Delta L$ of a bicrystal containing a (210)[001] STGB for various geometries, over time and over the number of absorbed vacancies.
The length change is observed to be strongly dependent on the area, with only a weak dependence on the total length.
The pore elimination time, roughly given by when $\Delta L$ becomes constant, is slightly dependent on total length.
A linear dependence of $\Delta L$ on the number of absorbed vacancies is also observed.}
\label{fig:geomvar_bi}
\end{center}
\end{figure}

Next, the influence of grain geometry on the length change is investigated in a bicrystal.
In \cref{fig:geomvar_bi} the results are shown for various grain lengths and areas.
In this and the following plots, only every 25th point is marked unless mentioned otherwise; the line always indicates all data points.
It is easily observed that the results are clustered by the grain area, whereas the grain length has no consistent influence on the results.
The small influence of the grain length is likely due to excess stress caused by finite size effects.
This also causes the pore to vanish at different times.
Note that there is a linear relationship between vacancy absorption and displacement.
By only plotting the system response over the number of absorbed vacancies, differences in kinetics can be removed from the problem.

\begin{figure}
 \includegraphics[width=\columnwidth]{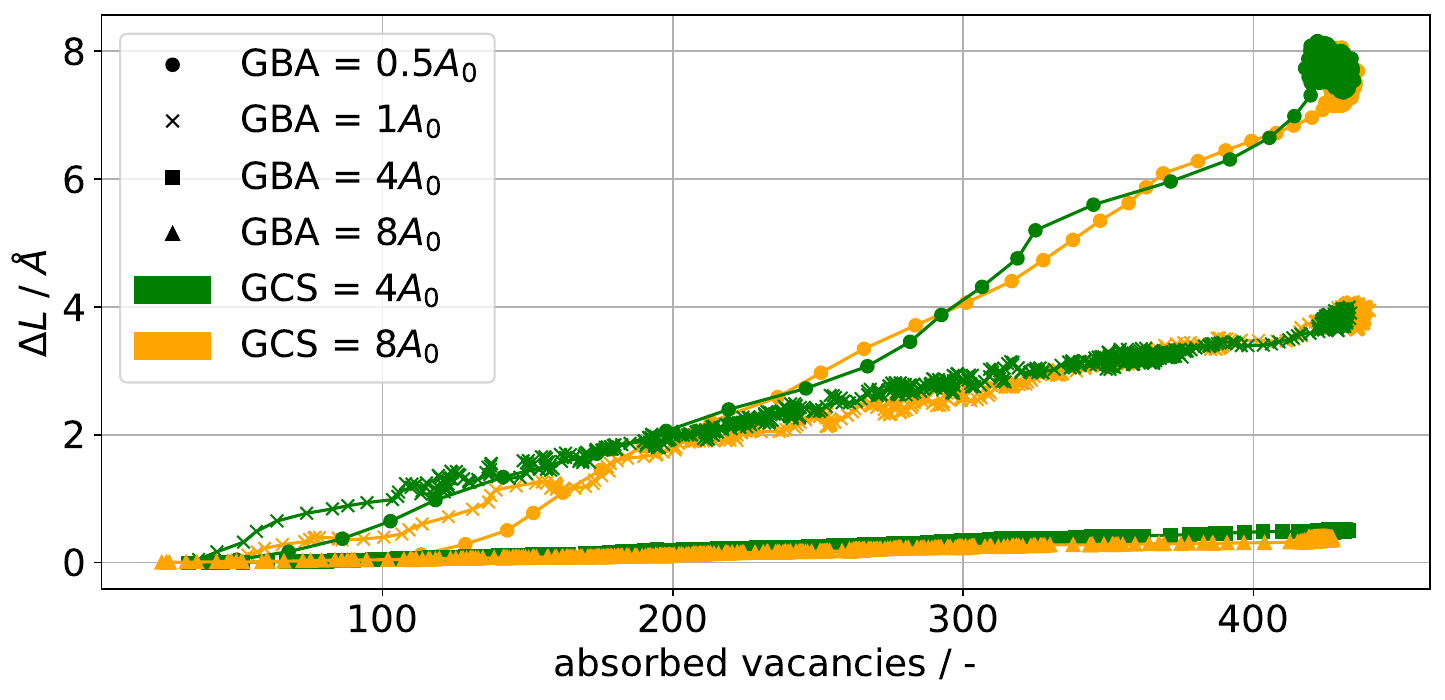}
 \caption{The length change $\Delta L$ of a bicrystal containing a (210)[001] STGB for two grain cross sections (GCS) and four grain boundary areas (GBA).
 Once GBA is independent of GCS, it is the determining factor of $\Delta L$.
 Due to the quick pore removal for GBA $=0.5A_0$ every point is marked in this plot.}
 \label{fig:geom-gbarea}
\end{figure}

As the present setup does not allow the differentiation of the GB area (GBA) from the grain cross section (GCS), a second series of simulations is performed.
In these the area around the grain boundary is largely removed, effectively decoupling the GB area from the grain cross section.
This is achieved by leaving only a rectangular area of size $\mathrm{GBA}$ in a region of length $\SI{50.61}{\angstrom}$ around the grain boundary.
Thus new surfaces are introduced to a thin region between the grains, which need to be relaxed.
This relaxation is done for \SI{16}{ns} before a spherical pore is placed on the GB, followed by another run for up to \SI{16}{ns}.
It should be noted that the newly generated surfaces can also act as vacancy sink/source, whose contribution to the supposed absorbed vacancy count is not easily accounted for.
Thus the relationship between displacement and vacancies shown in  \cref{fig:geom-gbarea} will differ from that of the previous results.
More specifically, more vacancies are being absorbed than plotted which also causes a larger displacement\footnote{Using the later model \cref{eq:model} and presuming that the volume change is caused by both area and length changes $\Delta V = A\Delta L + L\Delta A$ yields a bit more than twice as many vacancies being absorbed, which fits quite well with these results.}.
The squares and triangles show results for which $\mathrm{GCS}=\mathrm{GBA}$, whereas those in which $\mathrm{GCS}\neq\mathrm{GBA}$ are marked with crosses and dots.
The length change seems to be mainly influenced by GBA instead of GCS.

Hence we may formulate the first two rules of motion:
The length change of a bicrystal sample due to vacancy absorption is \emph{antiproportional} to the grain boundary area of the sample.
The length change of a sample due to vacancy absorption is \emph{proportional} to the number of absorbed vacancies.

\begin{figure}
\centering
 \includegraphics[width=\textwidth]{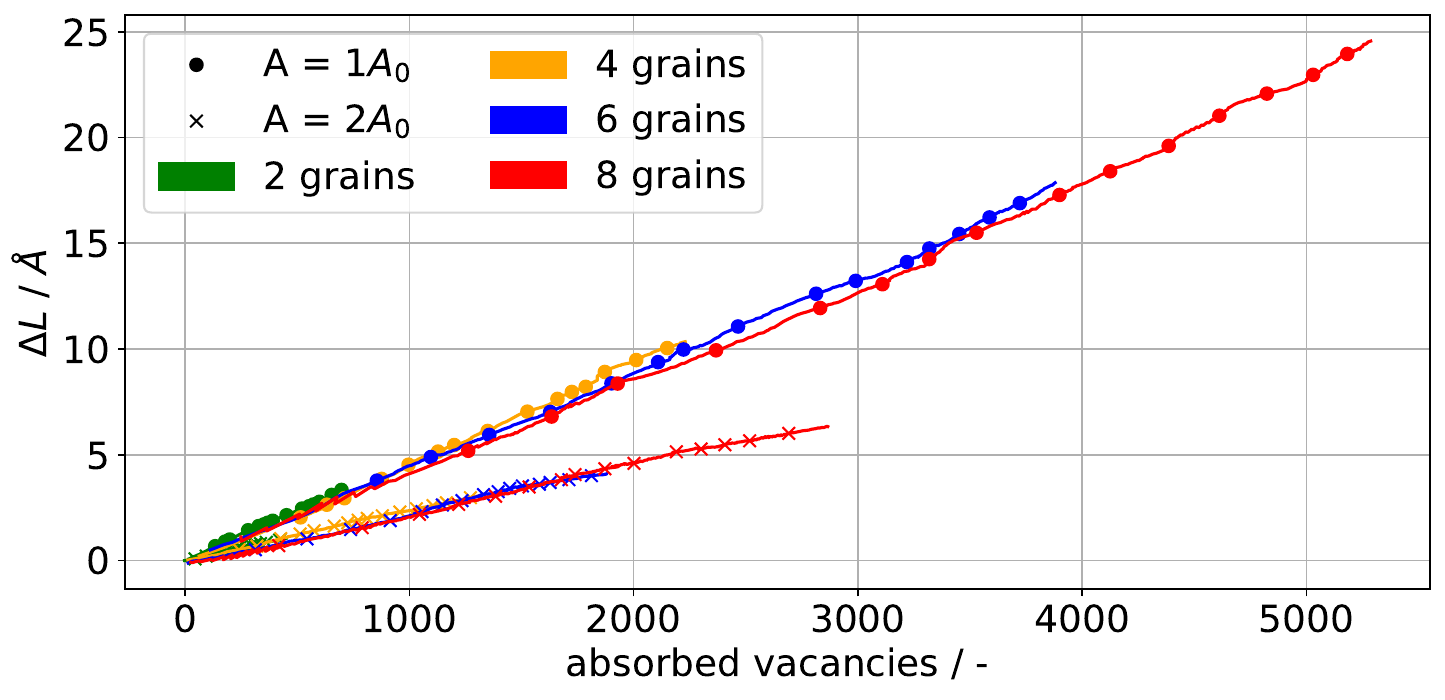}
 \caption{Total length change for two areas and up to 8 grains / 7 pores in the chain, with $L=2L_0$.
 A linear relationship between vacancies absorbed and the length change is observed.
 The results for longer chains roughly behave as if lying on the same line as for the smaller chains.}
 \label{fig:mdchain}
\end{figure}

\begin{figure}
\centering
 \includegraphics[width=\textwidth]{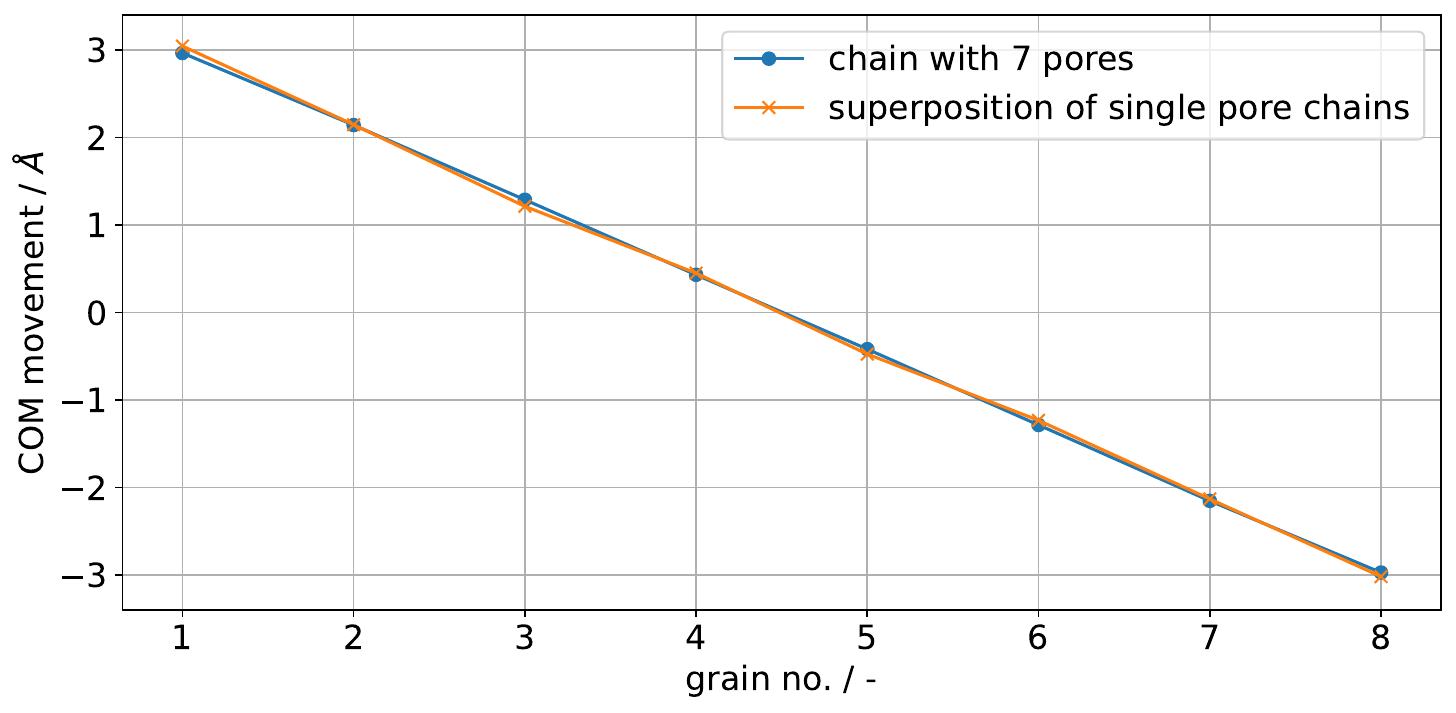}
 \caption{Comparison of a 8 grain chain with 7 pores and the solution via superposition.
 A good match for the total length change as well as the grain-specific displacements is observed.}
 \label{fig:superposition}
\end{figure}

Next we shall consider the influence of adding more grains, and hence grain boundaries and pores, to the system.
The results of investigating chains with up to 8 grains are depicted in \cref{fig:mdchain}.
For consistency, two grain boundary areas were considered and the first rule of motion is confirmed again.
If one follows the line described by a smaller simulation, one can then reasonably predict the length change observed in the larger chain.
Hence the displacement induced by the vacancy absorption on each grain boundary tends to be transported along the whole chain without any resistance.
This can be interpreted as a kind of superposition property inherent in the solution, i.e. for a system containing multiple vacancy absorption sites, the total solution is the sum of all single vacancy absorption site problems.
This is verified by running seven simulations for a chain with 8 grains, but with only a single pore being placed on a different grain boundary each time.
The seven individual solutions are then added together and compared against the full solution, shown in \cref{fig:superposition}.
The calculated solution and measured solution match well, hence we may define the third rule of motion:
The total length change of a system is determined by the superposition of all length changes due to vacancy absorption sites.

%


Recapping these rules we have the following properties:
\begin{enumerate}
 \item \label{itm:geometry} the length change is \emph{antiproportional} to the grain boundary area and \emph{independent} of the grain length
 \item \label{itm:vaccount} the length change is \emph{proportional} to the number of absorbed vacancies within the system
 \item \label{itm:superp} the length change of a system with multiple vacancy absorption sites is the \emph{superposition} of the individual length changes
\end{enumerate}

\begin{figure}
  \includegraphics[width=\textwidth]{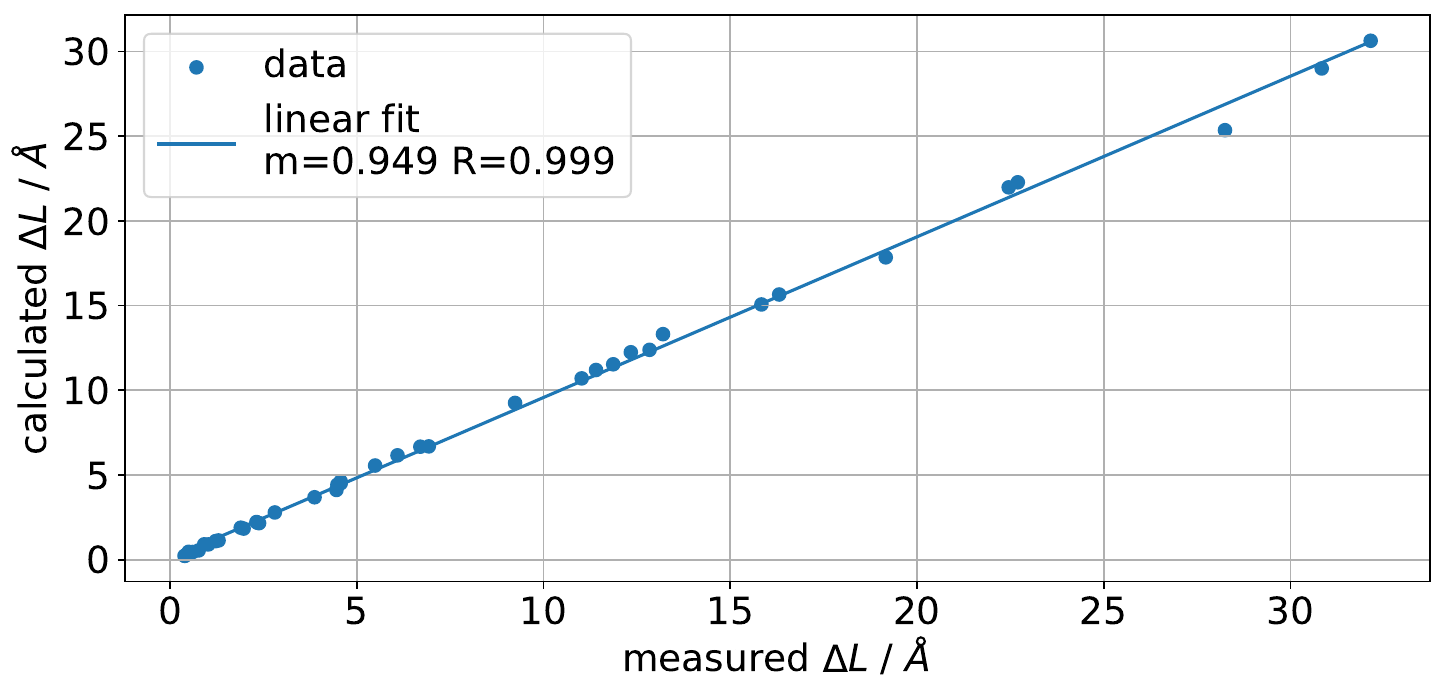}
  \caption{The measured length change is plotted against the calculated length change based on \cref{eq:model}.
  A general match is observed, with the fitted line's slope (m) indicating a slight underprediction ($5\%$) of the model.
  }
  \label{fig:model}
\end{figure}

In the following we shall shortly derive a model which contains all these properties.
In spirit it is rather similar to DeHoff's theoretical developments in the 1980s\cite{DeHoff1984,DeHoff1989}, but without requiring the grain structure to be decomposed into a space-filling cell structure.
First, based on \cref{itm:vaccount} we assume that each vacancy contributes a certain volume change $dV$ proportional to the atomic volume $\Omega$, with $dN$ being the number of vacancies which have just been absorbed.
Second, we assume that this volume change is due to a rigid movement of the entire crystal lattice of magnitude $dL$, i.e. $dV=AdL$ with the grain boundary area $A$, motivated by \cref{itm:geometry}
Thus one may write
\begin{align}
 \Omega dN &= AdL\\
 \leftrightarrow dL &= \frac{\Omega}{A} dN \label{eq:model-inc} \\ 
 \int \leftrightarrow \Delta L  &= \frac{\Omega}{A} \Delta N \label{eq:model}
\end{align}
with the length change $\Delta L$ taken to be positive for a shortening of the sample.
This model is verified by testing it not only against the already presented data, but also several grain boundary types, grain lengths, grain boundary areas, pore shapes, pore sizes, and numbers of pores.
The grain boundary area $A$ of each simulation is estimated based on the equilibrated area before the pore is placed.
The atomic volume $\Omega$ is determined by observing the volume of an fcc lattice of copper atoms in a periodic box at $T=\SI{700}{K}$ employing an NPT ensemble with $p=0$, resulting in $\Omega = \SI{1.22e-29}{m^3}$ which is close to the value given by \cite{Lawson2011}.
The number of absorbed vacancies is known via measurement and thus the length change can be determined.
The result of the comparison is shown in \cref{fig:model} which shows a good match for all data, though with a slight underprediction in the length change.

The presented model so far seems to only describe the total length change $\Delta L$ and not \corM{the motion of individual grains described by their} displacements $u_\alpha$, or equivalently velocities, as required in a field-resolved method as the phase-field method.
In order to resolve this, consider the implication of a length change $\Delta L$ in an effectively one-dimensional bicrystal $\alpha\beta$:
Since both grains move as rigid-bodies, the length change is the sum of the individual displacements and thus is given by 
\begin{align}
 u_\alpha - u_\beta &= \Delta L \\
 &= \Delta u_{\alpha\beta}
\end{align}
which is also the displacement jump $\Delta u_{\alpha\beta}$ across the grain boundary.
The superposition property \cref{itm:superp} is now exploited to enforce this simultaneously for all grain boundaries, leading to
\begin{align}
 \underbrace{\contact}_{B \times N }\underbrace{{u}}_{N} &= \underbrace{\Delta {u}}_{B} \label{eq:connectivity}
\end{align}
which is a usually overdetermined linear system of equations.
The matrix $\contact$ consists of rows with zeros and only one $+1$ and $-1$ each and acts on the unknown $N$ grain displacements $u = (u_1, \ldots, u_N)^T$.
The sign of the entries is determined by the one-sided grain boundary plane vector $n_\alpha = -n_\beta$ in the laboratory frame, with $n_\alpha$ being normal to the $\alpha\beta$ grain boundary and  pointing out of the $\alpha$ grain.
The right-hand side vector is determined by \cref{eq:model} for each of the $B$ grain boundaries.
This system may be solved e.g. in a least-squares sense, with the conservation of momentum accounted for afterwards by subtracting the mass-weighted average displacement from the solution ${u}$.
For the special case of a linear chain of grains, there are $B=N-1$ grain boundaries.
Adding conservation of momentum to the system of equations makes the matrix $\contact$ square and since the individual rows are linearly independent\footnote{A new unknown is introduced by each row and thus cannot be represented as a linear combination of previous rows. Conservation of momentum doesn't add a new unknown, but cannot be constructed from prior rows.}, it also is of full rank and thus uniquely solvable.
Furthermore, this formulation only accounts for motion due to vacancy absorption.
If other processes inducing displacement occur, e.g. grain boundary sliding, then these will require a separate treatment.

\begin{figure}
\begin{center}
 \begin{subfigure}[b]{0.45\textwidth}
 \includegraphics[width=\textwidth]{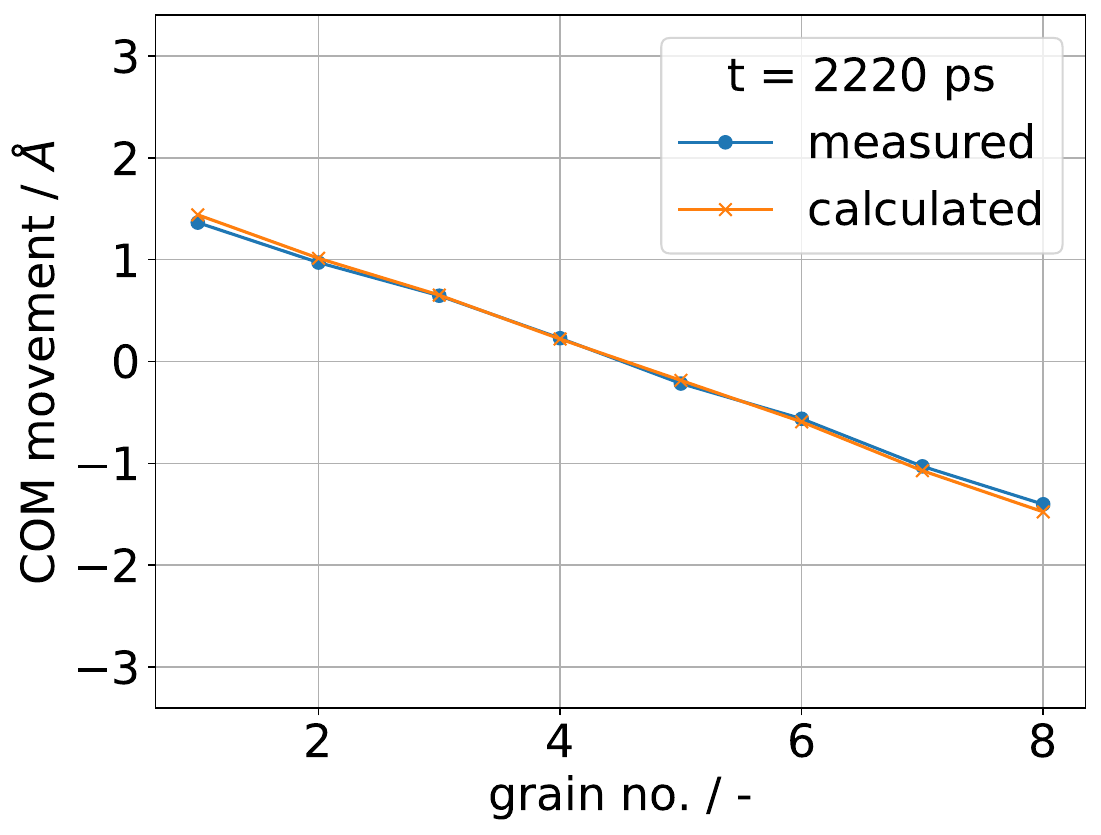}
 \end{subfigure}
\begin{subfigure}[b]{0.45\textwidth}
 \includegraphics[width=\textwidth]{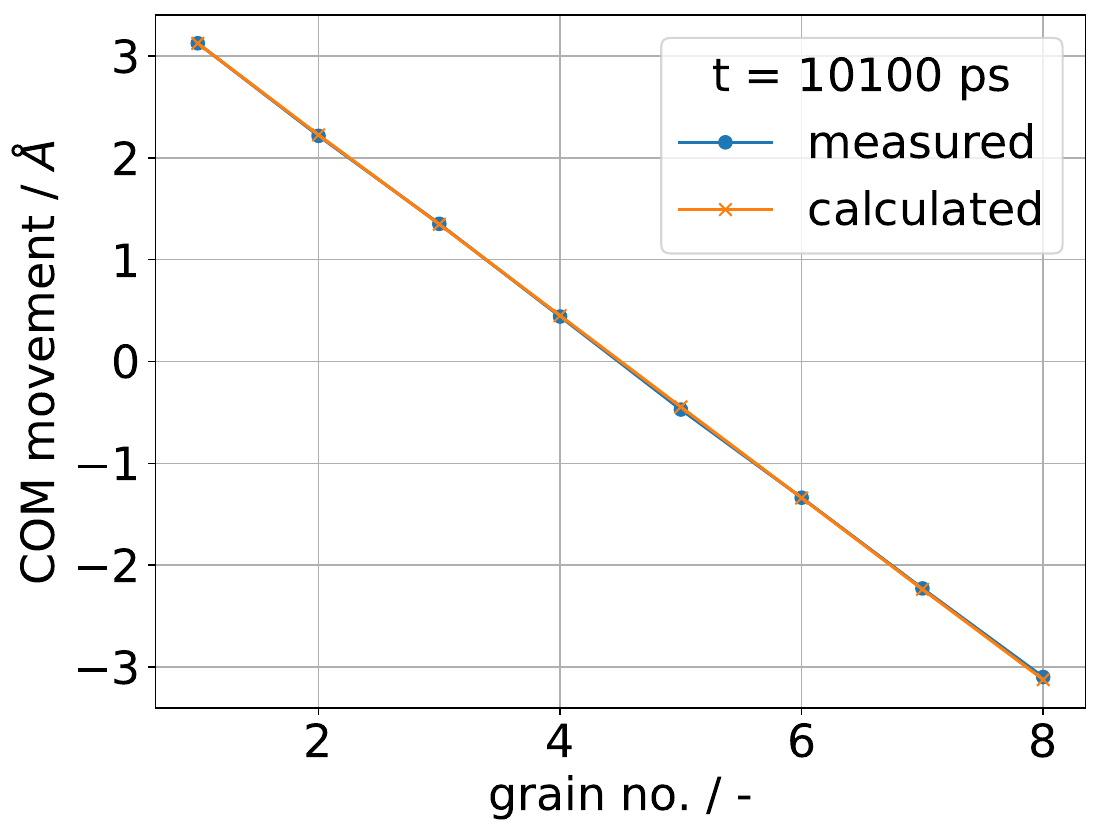}
 \end{subfigure}
 \caption{Comparison of solving \cref{eq:connectivity} and the observed data for two simulation states.
 A good match is observed for both.}
 \label{fig:results-syseq}
 \end{center}
\end{figure}

As a first test of this, we use the time-dependent data of one MD simulation and input these as the right-hand sides of the system of equations.
The contacts between grains which fill the matrix $\contact$ are also determined from these.
Since the chain is linear, including conservation of momentum makes it uniquely solvable.
The system is solved by direct matrix inversion since it is rather small, with a comparison of the calculated grain movement and the observed grain movement shown in \cref{fig:results-syseq} for two simulation states.
As the figure shows, there is a close match between the calculation and measurement, giving a measure of confidence in this approach.

\section{Phase-field}
In this section a new phase-field model will be described, following by a small-scale validation to ensure that the green body size effect is no longer present before large green bodies are calculated.

\subsection{Phase-field model with advection}
The model in the following is based on \cite{Hoetzer2018,Seiz2023a}, with the advection velocity being calculated with a model based on the MD results.
This new model in its entirety is dubbed MDi, as it is \emph{i}nspired by MD.
The evolution equations for the fields are the same as in \cite{Seiz2023a}:
\begin{align}
\pdiff{\phi_\alpha}{t} + \nabla \cdot (\vec{v}_\alpha \phi_\alpha) &= \frac{1}{\tau(\phi) \epsilon } \Big[
-\epsilon  \left(\pdiff{a(\phi,\nabla\phi)}{\phia} - \nabla \cdot \frac{\partial a(\phi,\nabla\phi)}{\partial\nabla\phia} \right) \label{eq:phi}  \\
&-\frac{1}{\epsilon}  \pdiff{w(\phi)}{\phia} \nonumber
- \sum^N_{\beta=0} \psi_{\beta}(\mu, T) \pdiff{h_{\beta}(\phi)}{\phia} \Big] - \lambda \nonumber,\\
\pdiff{\mu}{t} &=\left[ \sum_{\alpha=0}^N \ha(\phi) \left(\frac{\partial c_\alpha(\mu, T)}{\partial \mu} \right)\right]^{-1}  \nonumber \\
&\Biggl( \nabla \cdot \Big(M(\phi,\mu,T)\grad{\mu} - \vec{v}(x)c \Big) 
- \sum_{\alpha=0}^N c_\alpha(\mu,T) \frac{\partial h_\alpha(\phi)}{\partial t} \Biggr). \label{eq:mu} 
\end{align}
This represents the evolution of the $N$ phase-fields $\phi_\alpha$ and the chemical potential $\mu$ for one independent component, taken to be copper in the present paper.
The phase-field tuple $\phi$  distinguishes the surrounding vapour ($\phi_\Vap, \Vap = 0$) from copper grains of arbitrary orientation ($\phi_a, a > 0$).
The evolution of the chemical potential $\mu$ accounts for the species conservation via the concentration $c$ and takes into account the effect of phase changes due to changes in $\phi$.
For further particulars of the terms the interested reader is referred to \cite{Hoetzer2018,Seiz2023a}.

The calculation of the grain velocities follows the ideas outlined in the previous section.
This will be formulated in terms of instantaneous displacement $u$ and number density of absorbed vacancies $\Delta n$ to be consistent with the MD section.
\cor{The instantaneous velocity $v$ is linearly related to $u$ as $v = \frac{u}{\Delta t}$ with the time interval $\Delta t$.
The concentration $c$ is related to the number density $n$ as $n = \frac{N_a}{V_m} c$ with Avogadro's constant $N_a$ and the molar volume $V_m$.}
Each grain boundary $\alpha\beta$ absorbs an amount $\Delta N_{\alpha\beta} = \int_{GB} \Delta n_{\alpha\beta} dV$ of vacancies in a time interval $\Delta t$, with the density of absorbed vacancies $\Delta n_{\alpha\beta}$.
This is related to the vectorial displacement jump
\begin{align}
 V_{\alpha\beta} &=  \int_{GB} 4 \phi_\alpha \phi_\beta dV \\
 A_{\alpha\beta} &=  \frac{V_{\alpha\beta}}{l_0} \\
 \vec{\Delta u}_{\alpha\beta} &= \frac{1}{V_{\alpha\beta}} \int_{GB} 4 \phi_\alpha \phi_\beta \frac{\Omega}{A_{\alpha\beta}} \Delta n_{\alpha\beta}  \vec{n_{\alpha\beta}} dV \label{eq:dispjump} 
\end{align}
in which the orientation of the grain boundary plane was taken into account by a similar approach as Wang\cite{Wang2006}, but employing normalized phase-field gradients representing the normal vector, i.e. $\vec{n_{\alpha\beta}} = \frac{\nabla (\phi_\alpha-\phi_\beta)}{|\nabla (\phi_\alpha-\phi_\beta)|}$ instead of just the phase-field gradients.
Note that $\vec{n_{\alpha\beta}}$ needs to be chosen consistently with the later input to the matrix equation.
It is always taken to be from the lower $\alpha$ index to the larger $\beta$ index, defining a unique orientation for each $\alpha\beta$ pair.
The grain boundary region $GB$ is defined to be the region in which $g = \phi_\alpha \phi_\beta \geq g_T$ holds, i.e. only the region where both grain phases have significant volume fractions.
In the following the value $g_T = 0.14$ is chosen.
The grain boundary area  $A_{\alpha\beta}$ is resolved by dividing the grain boundary volume $V_{\alpha\beta}$ by the equilibrium grain boundary width $l_0 = \int_0^1 4 \phi_\alpha(x) (1-\phi_\alpha(x)) d\phi = \frac{\pi^{2} \epsilon}{8} $ for the employed obstacle potential.
The remaining  $4\phi_\alpha\phi_\beta$ and $V_{\alpha\beta}$ terms act as a weighted average to assign higher weight to regions which contain more grain boundary phase $4\phi_\alpha\phi_\beta$.

\Cref{eq:dispjump} still contains an unknown, namely the number of absorbed vacancies.
For this we assume that a grain boundary has a certain equilibrium number density of atoms $n^{gb}_{eq}$ and that it relaxes towards this number density:
\begin{align}
 \pdiff{n}{t} = -\frac{n-n^{gb}_{eq}}{t_r} \label{eq:vacabs}
\end{align}
which allows the calculation of the number density of absorbed vacancies $\Delta n =  - \pdiff{n}{t} \Delta t$.
\corM{Since this describes a relaxation process, one can consider the term $n-n^{gb}_{eq}$ as the driving force for this process --- once it vanishes, densification via advection stops.}
Note that this assumes atoms and vacancies are both conserved quantities.
Only the number of atoms is actually conserved, with vacancies and lattice sites being destroyed and generated during absorption to accommodate the volume change.
However, this rough treatment suffices to show the capabilities of the model and is in fact quite standard in phase-field modelling of sintering\cite{Wang2006,Biswas2016,Hoetzer2019,Greenquist2020,Seiz2023a}.
In spirit this is similar to Wang's model\cite{Wang2006} but the relaxation time $t_r$ is identified explicitly here, which can in turn be determined via molecular dynamics.
In the MD simulations $t_r$ was observed to depend strongly on the grain boundary orientation relationship; it is strongly related to the efficiency of a grain boundary at absorbing vacancies such as described in \cite{Uberuaga2015}.
In the following simulations, all grain boundaries are assumed to be of a (210)/[001] STGB ($\SI{53.1}{\degree}$) type.
Furthermore, $n^{gb}_{eq}$ does not have the meaning of a ``grain boundary equilibrium density'' in this context.
Specifically, if it is below the bulk density as one would expect based on physics, it will \emph{push apart} grains instead of attracting them.
This phenomenon has been investigated in-depth in \cite{Seiz2022,Seiz2023a}.
Based on the suggestions therein, $n^{gb}_{eq}$ is calculated based on the observed average chemical potential on the particle surface $\hat{\mu}$, which should approximate the capillary pressure.
This ensures that the resulting velocities are consistent with the free energy functional and the theoretical dihedral angle is recovered\cite{Seiz2023a}.
Note that this allows a rather trivial addition of an external isotropic pressure to the system, as the capillary pressure can simply be shifted by the external pressure.

It should also be noted that properties such as $\vec{\Delta u}_{\alpha\beta}$ and $V_{\alpha\beta}$ need to be tracked for each grain boundary individually and thus their memory and communication requirements scale as $O(nN^2)$ if implemented na\"{i}vely, with the number of parallel processes $n$ and number of phase-fields $N$.
While this is not a problem for a few hundreds of grains, once thousands or tens of thousands grains are resolved this will dominate the memory and communication costs.
This is resolved by only storing the actual contacts (thus being a sparse representation) and distributing it across all parallel processes.
The message passing interface (MPI) is employed for the parallelization and updates to this distributed matrix are realized via one-sided communication.
The details of this scheme will be published in a separate paper.

\begin{figure}
  \centering
  \includegraphics[width=0.5\textwidth]{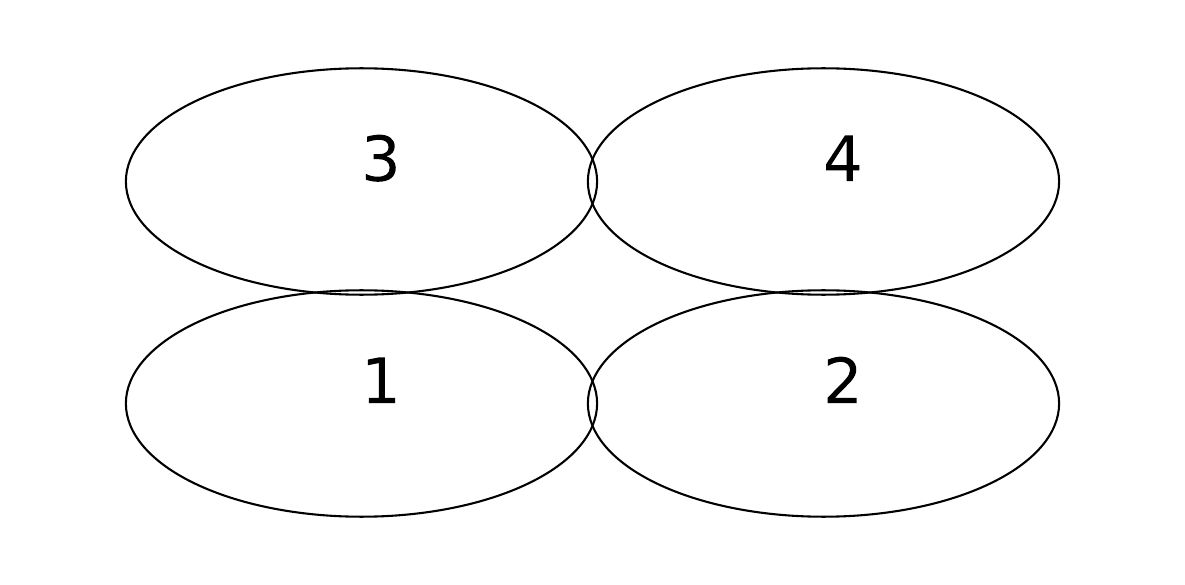}
  \caption{Example 2x2 setup of grains for clarifying the matrix structure.
  The number within the circles indicates the grain index.}
  \label{fig:2x2}
\end{figure}

The displacement jumps are used to solve for the particle displacements $u$, for each direction separately, by building a system of equations
\begin{align}
 \contact{} u &= \Delta u \label{eq:contact}
\end{align}
in which the matrix $\contact$ is filled according to the connectivity determined during the simulation.
The structure of the matrix is clarified by the following example:
Consider a 2x2 grid of grains, depicted in \cref{fig:2x2}.
Each particle has two contacts, one along each dimension.
These are always taken to be from the lower grain index $a$ of $\phi_a$ to the higher one, i.e. we have the contacts described by the ordered set $\{ C_{1,2}, C_{1,3}, C_{2,4}, C_{3,4} \}$.
The size of this set is equal to the number of grain boundaries $B$ in the system.
A matrix $C_d$ of dimension $B \times N $ is constructed per dimenions $d$, with a corresponding RHS $\Delta u_d$ of size $B$.
For each contact a row is added to the matrix, with only non-zero entries for those grains which are connected by this contact.
The magnitude of the entries is always 1, but the sign is determined consistently with $\vec{n_{\alpha\beta}}$ in this direction, going consistently from the lower to the higher index.
The right-hand side displacement jump is already in a vector form and therefore can be easily split into its components $\Delta u_d$.
Thus \cref{eq:contact} can be written, for the $x$ dimension, as
$$
\begin{pmatrix}
 1 & -1 & 0 & 0 \\
 -1 & 0 & 1 & 0 \\
 0 & -1 & 0 & 1 \\
 0 & 0 & 1 & -1 \\
\end{pmatrix}
\begin{pmatrix}
 u_{x,1}\\
 u_{x,2}\\
 u_{x,3}\\
 u_{x,4}\\
\end{pmatrix}
= \begin{pmatrix}
\Delta u_{x,1,2}\\
\Delta u_{x,1,3}\\
\Delta u_{x,2,4}\\
\Delta u_{x,3,4}\\
\end{pmatrix}
$$
with the sign on the left-hand side determined based on the grain boundary normal.
If the grain boundary normal has no component in a dimension, the sign within the matrix plays no role as the right-hand side will be zero.
This effectively says that no relative motion occurs.
Note that this matrix, although square, is singular, since e.g. the fourth row can be constructed by adding the first and second row and subtracting the result from the third row.
Conservation of momentum is accounted for afterwards by subtracting the mass-weighted average displacement from each grain displacement.

The resulting system of equations is usually overdetermined and only in the special case of a linear chain of particles can be solved exactly if conservation of momentum is included in the system.
However, the system can be solved in a least-square sense.
This could be done with e.g. \corM{an Alternating Direction Method of Multipliers} approach\cite{Boyd2010}, which requires a collective reduction of $N$ scalars per iteration.
Alternatively, the matrix $\contact{}$ and its transpose could be partially distributed and then be solved by using some Krylov subspace method (e.g. a block \corM{Conjugated Gradient Least Squares} method\cite{Ji2017}) which would need a collective reduction of a single scalar per iteration.
However, both approaches lead to excessive communication time (on the order of entire field sweeps) and thus an approach suited to the present problem is developed.

Typical solutions of the system were sought by generating packings, from which \cref{eq:contact} was determined while assuming the right-hand side is given by a normal distribution
\begin{align}
 p(\xi) &= \frac{1}{\sqrt{2\pi\sigma^2}} \exp(-\frac{(\xi-\mu)^2}{2\sigma^2}) \label{eq:gaussian}
\end{align}
with its mean $µ$ and standard deviation $\sigma$.
A mean of $\mu=1$ is generally employed with variable standard deviation $\sigma$.
A spatial dependence can be included by simply adding a function of particle position to the random sample generated by \cref{eq:gaussian}.
The system is solved employing the \corM{Least Squares via QR factorization (LSQR)}\cite{Paige1982} algorithm.

It was observed that the spatial distribution of the right-hand side tends to determine the shape of the solution.
If it is assigned randomly without any spatial dependence, a mostly linear function of position is observed, with local inhomogeneities.
If a linear dependence on the position is added, the displacement field becomes a quadratic function of position.
\corM{This implies that the solution of the system basically integrates the spatial distribution of right-hand sides.
This can also be seen from the structure of the matrix $\contact{}$:
Each row effectively represents a finite difference formula, with the right-hand side giving the slope, i.e. $\contact{}$ is a differentiation operator.
Thus its generalized inverse is an integration operator.}

Given that the same grain boundary type is assumed for all contacts, with similar initial neighbourhoods, it seems reasonable to assume the displacement field is given by a linear function.
Hence one may approximate the full problem by replacing the particle displacements $u$ by the relation
\begin{align}
 u = m (x-x_m) \label{eq:linearfunc}
\end{align}
with the known center of mass of each grain $x$, the known total center of mass $x_m$ and an unknown slope $m$.
Hence only $m$ remains to be determined, which can be done exactly in a least-square sense by employing the normal equations:
\begin{align}
 \underbrace{\contact}_{C \in \mathbb{R}^{BxN}} u = \underbrace{\contact (x-x_m)}_{D \in \mathbb{R}^{B}} m &= \Delta u \\
 Dm &= \Delta u \\
 \underbrace{D^TD}_{q \in \mathbb{R}} m &= \underbrace{D^T \Delta u}_{p \in \mathbb{R}}\\
 m &= \frac{p}{q}
\end{align}
which only requires a parallel reduction operation for $p$ and $q$.

\begin{figure}[]
\centering
     \begin{subfigure}[]{\columnwidth}
        \includegraphics[width=\textwidth]{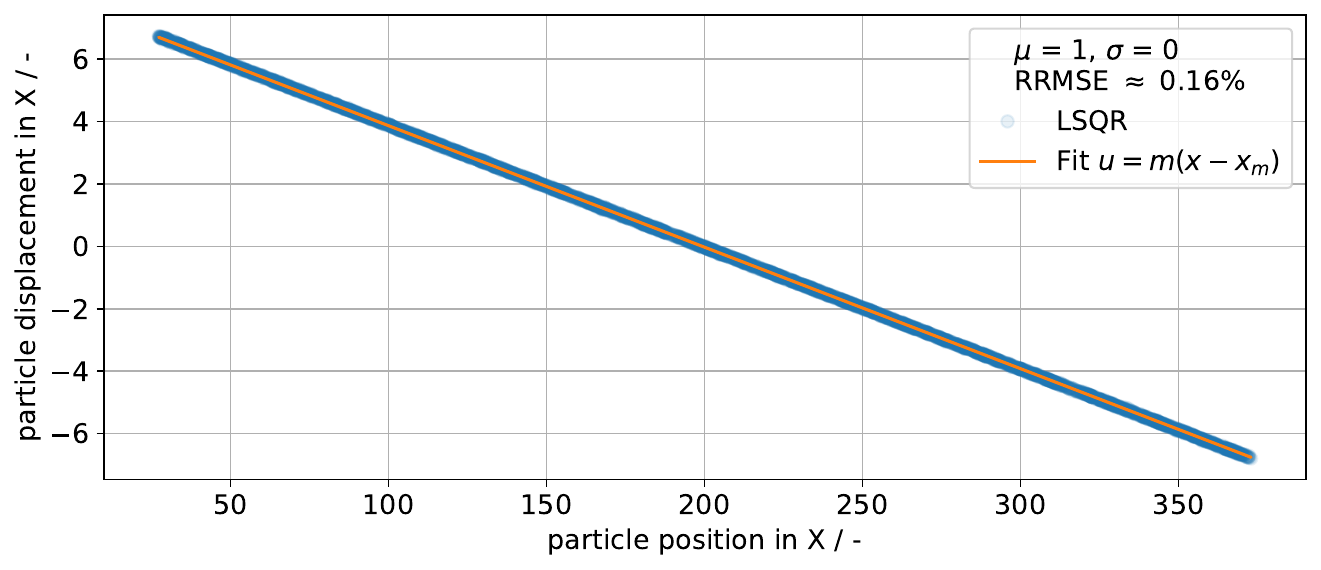}
    \end{subfigure}
~
    \begin{subfigure}[]{\columnwidth}
    \centering
        \includegraphics[width=\textwidth]{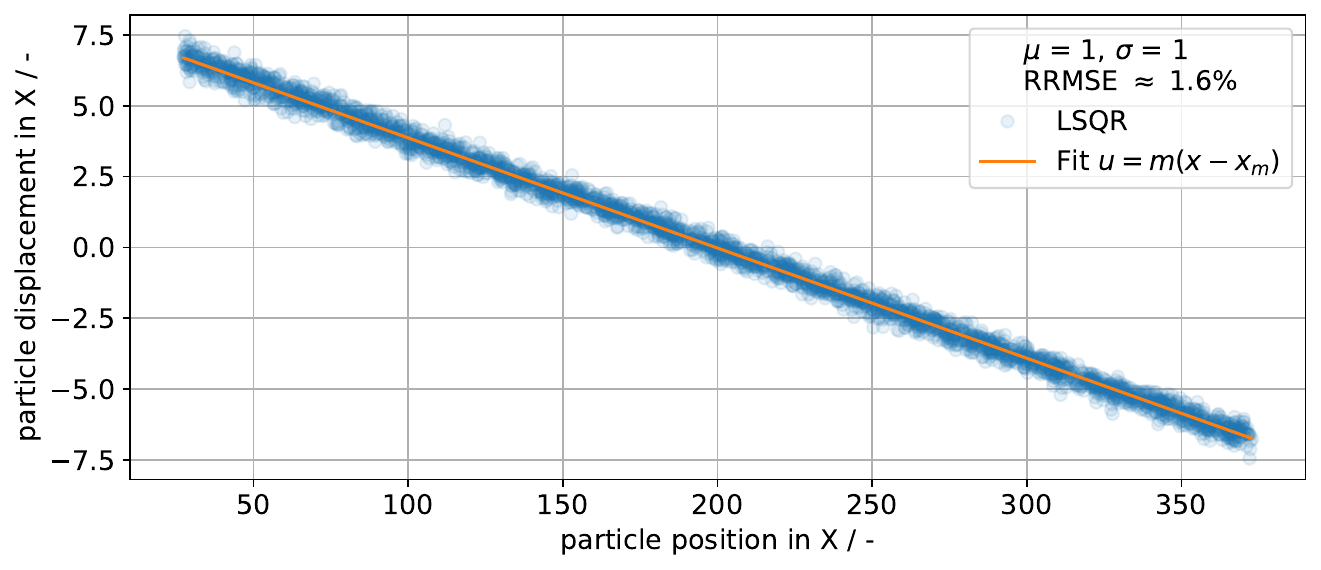}
    \end{subfigure}
 \caption{Comparison of the full displacement solution calculated via LSQR and fit ansatz for two values of the standard deviation.
 Each marker indicates a particle's position and its resulting displacement in one dimension of the 3D packing.
 The shape and scale of the solution are always well-preserved by the fit ansatz.
 }
 \label{fig:fitcomparison}
\end{figure}

This approach is compared against the full solution via LSQR.
The error is evaluated with the root-mean-square error (RMSE) and relative RMSE (RRMSE) defined as
\begin{align}
 \mathrm{RMSE} &= \sqrt{\frac{1}{N} \sum_i (x_{i,LSQR} - x_{i,lin})^2} \\
 \mathrm{RRMSE} &= \frac{\mathrm{RMSE}}{\max(x_{LSQR}) - \min(x_{LSQR})}
\end{align}
for the solution vectors of the LSQR method and the linear fit ansatz.
It is generally observed that the RRMSE is unaffected by the choice of mean $\mu$, while the RMSE changes due to the scale in displacement.
For regular packings the (R)RMSE is observed to scale mostly linearly in the standard deviation $\sigma$ of the random distribution, with errors on the order of machine precision for $\sigma = 0$.
The irregular packings employed later for the PF simulations generally show some non-zero error even for $\sigma = 0$, though starting from about $\mathrm{RRMSE} \approx 0.16\%$.
Even if $\sigma$ is comparable to the mean of the normal distribution, $\mathrm{RRMSE} \approx 1.6\%$ and thus still quite acceptable.
A visual comparison for the effect of $\sigma$ on the displacements in a 3D packing containing 3445 particles is shown in \cref{fig:fitcomparison}.
As can be seen, the solution shape and scale are always well-preserved.
What the fit ansatz obviously cannnot match however is the local variation of absorption activity, modelled by the random distribution of displacement jumps.
The interested reader is referred to the Supplementary Material wherein the code employed for this test is published in full, along with the irregular packings employed later for the large-scale sintering simulations.

As a final note, the specification of the vacancy absorption rate is the main weakness of the presented model, since it cannot be completely linked to quantities obtainable via MD. 
\corM{Hence a direct comparison to MD simulations would likely yield a significant mismatch in the temporal evolution.}
However, if an improved model for the vacancy absorption rate is developed, it can be included easily into the current approach.
This is due to the model effectively being split into a kinematic part \cref{eq:contact} and a dynamic part \cref{eq:vacabs} which can be changed independently.


\subsection{Parameters and data evaluation}
The scales employed are listed in \cref{tab:nondim} and the materials parameters in \cref{tab:params}.
These are the same as in \cite{Seiz2023a} except for the newly introduced atomic volume $\Omega$ and the relaxation time $t_r$.
The relaxation time is determined by running MD simulations for a (210)/[001] STGB ($\SI{53.1}{\degree}$) in which atoms are randomly removed from the grain boundary and observing the time it takes until the atom count within the grain boundary has stabilized.
Several such simulations were run and the order of magnitude for the relaxation time then used for the value of $t_r$.
\corM{
This is more of qualitative approach, but $t_r$ behaves similarly to the stiffness $\kappa$ in the classical rigid-body motion (RBM) model of Wang\cite{Wang2006}:
In the classical model, the RBM velocity scales as
\begin{align}
 v \propto \kappa (n-n_{gb}^{eq})
\end{align}
whereas in the present model it is rather
\begin{align}
 v \propto  \frac{n-n_{gb}^{eq}}{t_r}
\end{align}
i.e. the velocity is proportional to $\kappa$, but inversely proportional to $t_r$.
Shi et al.\cite{Shi2021} could show that once $\kappa$ is sufficiently large, the resulting advection velocity does not change upon a further increase in $\kappa$.
Hence the proportionality is only applicable to a certain limit, after which other processes control the velocity.
Since $t_r$ divides the driving force $n-n_{gb}^{eq}$ whereas $\kappa$ multiplies it, the same behaviour applies, but reversed:
Once $t_r$ is small enough, making it smaller will not change the velocity.
One can think of this as saying that the problem ought not to be controlled by the absorption rate and hence the absorption rate should be appreciably faster than the slowest other process.}
The time step is determined by calculating stable time step widths within the explicit scheme and employing the minimum to ensure stable time integration as described in \cite{Seiz2023a}, with the table only listing the largest stable timestep with zero advection velocity.

\corM{The evaluation of the data is the same as in \cite{Seiz2023a} in terms of strain and density.
The contact number later employed in the three-dimensional simulations is calculated with the package cc3d's\cite{cc3d} function \texttt{contacts} on the phase-field label field and the simply counting the pairwise occurrences of labels.
The surrounding vapour phase is treated as background, with the phase-field label field being defined pointwise as the label which has the highest phase-field value.
Since surface particles always have some missing neighbours, including these would induce a particle size and simulation size bias in the coordination number.
These are excluded by determining the bounding box of the green body, then shrinking it by \SI{32}{nm} in each direction and only averaging over particles contained in this shrunken box.
In the 3D simulation this excludes at least two particle layers, removing the surface effect.}

\corM{The simulations in \cref{sec:vali} are conducted locally while employing GNU Parallel\cite{parallel} for efficient job management.
The three-dimensional simulations in \cref{sec:3d} are calculated on the Hawk supercomputer at the High Performance Computing Center in Stuttgart.
Hence the processor employed for the three-dimensional simulations is the AMD EPYC 7742, with 64 cores running at \SI{2.25}{GHz}.
A single core performance of \SI{9.4}{GFLOP/s}, i.e. $\SI{26.1}{\percent}$ of the theoretical peak, is achieved.}

\begin{table}[h]
\centering
\caption{nondimensionalization parameters}
\label{tab:nondim}
 \begin{tabular}{ll}
scale   & value \\
length $l_0$ &  $\SI{1e-8}{m}$  \\
diffusivity $D_0$ & $\SI{1e-12}{m^2.s^{-1}}$\\
time $t_0$ & $\SI{1e-4}{s}$ \\
velocity $v_0$ & $\SI{1e-4}{m.s^{-1}}$\\
temperature $T_0$ & $\SI{700}{K}$ \\
surface energy $E_{s,0}$ & \SI{1}{J.m^{-2}} \\
energy density $E_{r,0}$ & $\SI{1e8}{J.m^{-3}}$ \\
molar volume $V_{m,0}$ & $\SI{7.1e-6}{m^3.mol^{-1}}$ \\
\end{tabular}
\end{table}

\begin{table}[h]
\centering
\caption{Employed physical and numerical parameters for the simulations.}
\label{tab:params}
 \begin{tabular}{lll}

parameter   & nondim. value & physical value \\
\multicolumn{3}{c}{\textit{numerical parameters}}\\
grid spacing $\Delta x$   &    0.1  &   $\SI{1e-9}{m}$  \\
max. time step $\Delta t_{max}$   &    $\num{1.5e-05}$  & $\SI{1.5e-9}{s}$  \\
interface parameter $\epsilon$   &    $4\Delta x$  &   $\SI{4e-9}{m}$  \\
interface width $W\approx2.5\epsilon$   &    $10\Delta x$  &   $\SI{10e-9}{m}$  \\
grain boundary cutoff $\phi^{min}_{\alpha\beta}$ & 0.14 & - \\
\multicolumn{3}{c}{\textit{physical parameters}}\\
surface energy $\gamma_{v\alpha}$     & 2 & $\SI{2}{J.m^{-2}}$ \\
grain boundary energy $\gamma_{\alpha\beta}$     & 1 & $\SI{1}{J.m^{-2}}$ \\
volume diffusion $D$    & \num{1e-3}               & $\SI{1e-15}{m^2.s^{-1}}$  \\
grain boundary diffusion  $D_{gb}$    & 55               & $\SI{5.5e-11}{m^2.s^{-1}}$  \\
surface diffusion  $D_s$ & 169               & $\SI{1.69e-10}{m^2.s^{-1}}$   \\
physical interface width $\delta_i$ & 0.02 & $\SI{2e-10}{m}$ \\
surface kinetic coefficient $\tau_{\Vap\alpha}$ & 0.08 & $\SI{8e10}{J.s.m^{-4}}$ \\
grain boundary kinetic coefficient $\tau_{\alpha\beta}$       & 100 $\tau_{\Vap\alpha}$ & $\SI{8e12}{J.s.m^{-4}}$ \\
effective stiffness $\kappa$ & 3200 & - \\
atomic volume $\Omega$ & $\num{1.22e-5}$ & $\SI{1.22e-29}{m^3}$\\
GB relaxation time $t_r$ & $\num{1e-8}$ & $\SI{1e-12}{s}$
\end{tabular}
\end{table}

\subsection{Validation}
\label{sec:vali}
\corM{
In this section the system size convergence of the present model will be investigated.
In effect, this tests whether the superposition rule of motion from the MD simulations has been transferred successfully.
As will be shown below, the results for a two-particle model are virtually identical between the present model MDi and a grand potential model including advection, i.e. the model (ADV-$\mu$) of \cite{Seiz2023a}.
Hence the accordance with classical theory in terms of neck growth and approach of centers, as well as Herring's scaling laws for a two-particle model as shown in\cite{Seiz2023a}, are transferable to the present model.
Thus the strain in a system of increasing size will be investigated, first in a particle chain as suggested by \cite{Seiz2022}, then in a rectangular grain geometry.
Simulations are run for both models, with the number of particles in the chain given by $n \in \{2, 4, 8, 16, 32\}$.}
The simulations for the same geometry are all run to the same simulation time $t_e$.
The strain is evaluated based on the movement of the barycenters of the first and last particle, i.e. $e = \frac{L(t)-L(0)}{L(0)}$ with $L(t) = x_{m,last}-x_{m,first}$ and $x_{m,*}$ being the the barycenter of the first/last particle.
\corM{The geometries considered in this section can also in general be used as a relatively cheap benchmark geometry for determining whether size-independent densification is captured by the model.}

\begin{figure}[]
\begin{center}
  \begin{subfigure}[]{0.9\columnwidth}
  \centering
        \includegraphics[width=\textwidth]{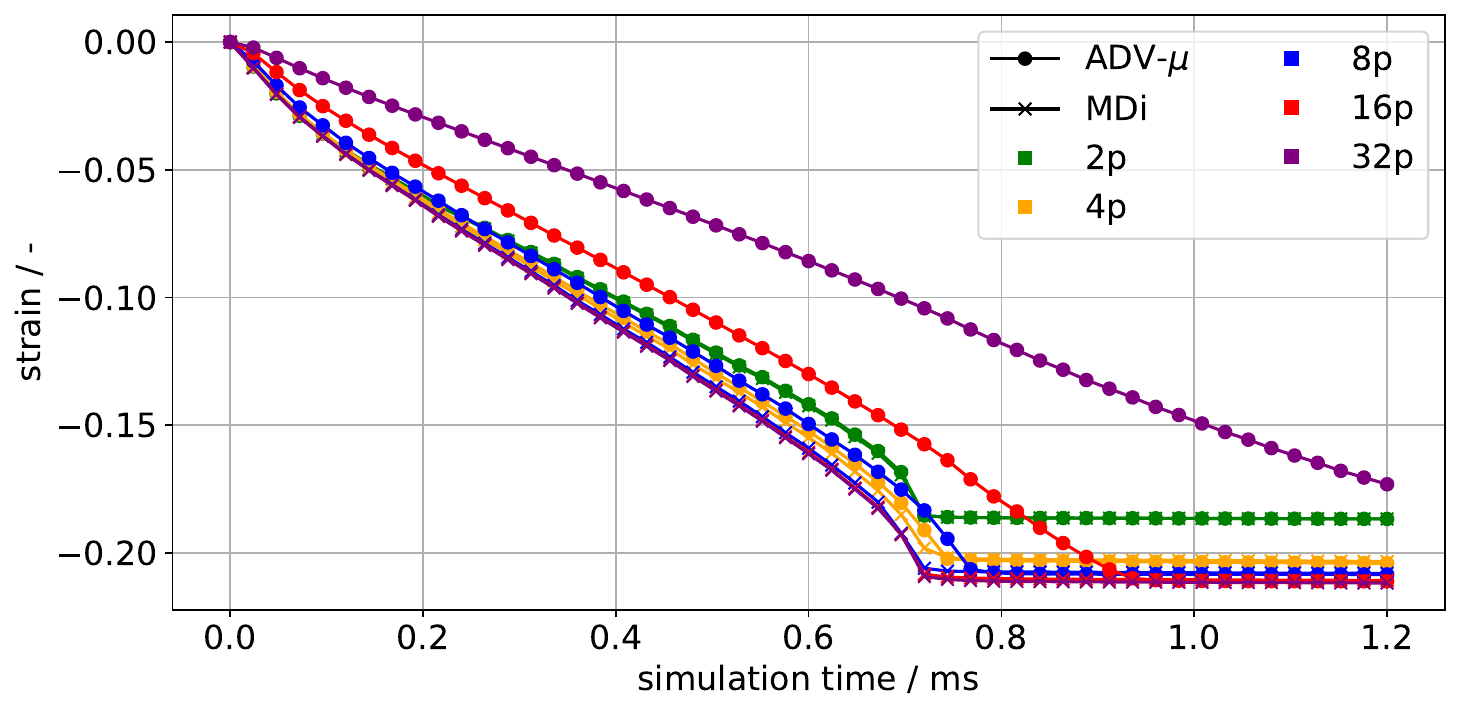}
        \caption{strain over time}\vspace{0.25cm}
 \label{fig:strain-wang-md}
    \end{subfigure}
  \begin{subfigure}[]{0.9\columnwidth}
    \centering
        \includegraphics[width=\textwidth]{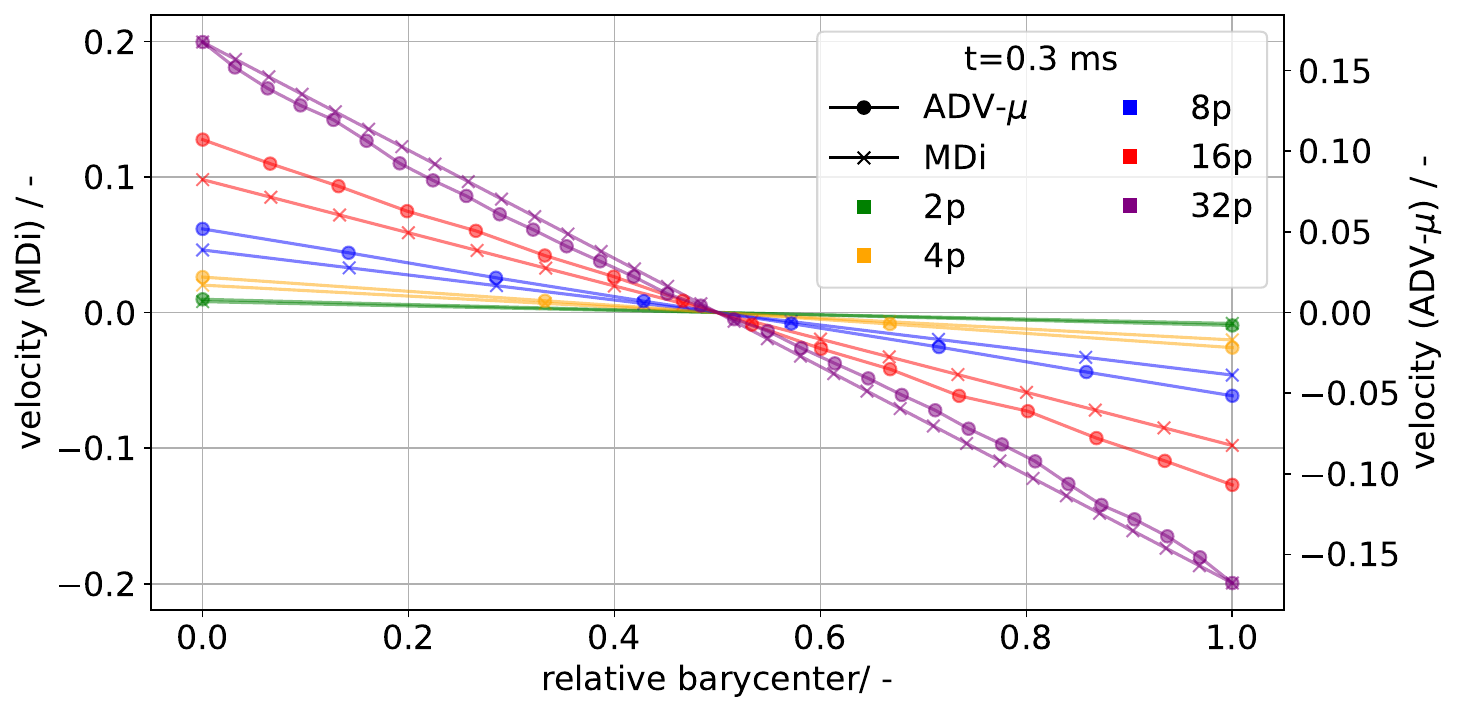}
        \caption{velocity distribution}
        \label{fig:velo-wang-md}
    \end{subfigure}
 \caption{Comparison of the MDi model (crosses) and the model of \cite{Seiz2023a} (circles).
 The dependence on the strain becomes negligible at around 16 particles for the MDi model but is substantial across all investigated particle counts for model ADV-$\mu$.
 The velocity distribution shows the inhomogeneous densification (non-constant velocity gradient between the particles) from which this lack of convergence originates.
 }
 
 \label{fig:dens}
 \end{center}
\end{figure}

The strain over time for a chain of circular particles is shown in \cref{fig:strain-wang-md}.
While model MDi seems to converge at around 16 particles in a chain, model ADV-$\mu$ still shows a large change at this particle count.
Let us first consider how model ADV-$\mu$ fails to converge by looking at the velocity distribution in \cref{fig:velo-wang-md}.
The velocity is plotted over the relative barycenter $X_{i,r} = \frac{x_{m,i}-x_{m,first}}{x_{m,last}-x_{m,first}}$ which allows the compact viewing of chains of arbitrary physical length.
The absolute scale of velocity reached is similar for both models, but while model MDi always yields a linear function by design, model ADV-$\mu$ tends to produce curved velocity profiles which directly imply inhomogeneous densification.
While a small amount of inhomogeneity is to be expected due to the discussion below, this should drop off rapidly from the outermost particle.

\begin{figure}
\centering
\includegraphics[width=\columnwidth]{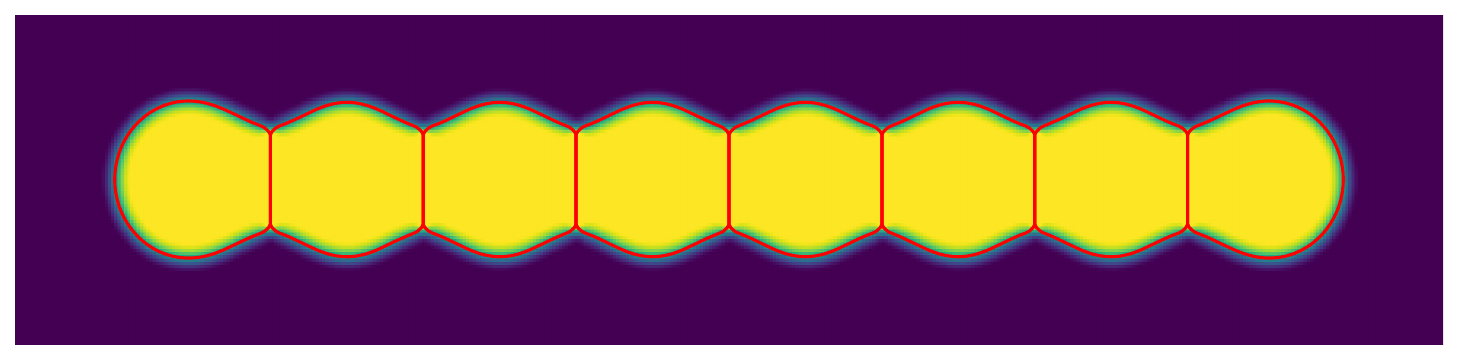}
 \caption{The grain field, defined by $1-\phi_\Vap$, at simulation end for 8 particles and model MDi.
 Yellow indicates the copper grains, dark indigo the surrounding vapour and green the interface between them.
 The phase transitions defined by the 0.5-isoline of the phases are represented with red lines.}
 \label{fig:concfields}
\end{figure}

One might ask why model MDi does not instantly converge in this case, given that the model without the phase-field information did so?
This is due to the chain not actually being homogeneous and only becoming so at a sufficient number of particles.
The simulation geometry at the end of the simulation is depicted in \cref{fig:concfields}, showing that the chain ends take up a different shape than inner chain particles.
This shape difference is simply due to the inner particles being restricted from free movement, whereas the outer particles can easily adjust.
This difference will also eventually lead to grain growth.

\begin{figure}[]
\begin{center}
     \begin{subfigure}[]{0.9\columnwidth}
     \centering
        \includegraphics[width=\textwidth]{images_strain-wang-md-block}
        \caption{strain over time}\vspace{0.25cm}
 \label{fig:strain-wang-md-block}
    \end{subfigure}
    \begin{subfigure}[]{0.9\columnwidth}
    \centering
        \includegraphics[width=\textwidth]{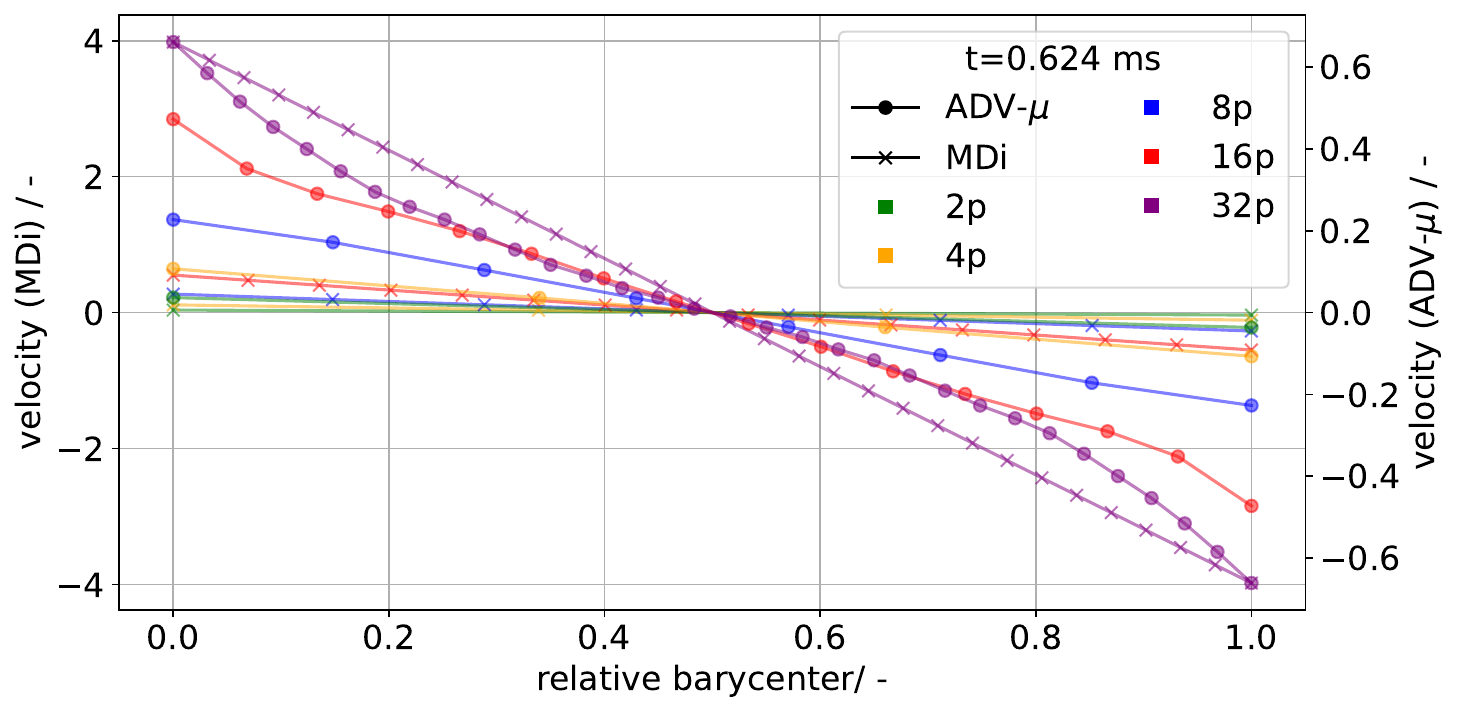}
        \caption{velocity distribution}
        \label{fig:velo-wang-md-block}
    \end{subfigure}
 \caption{Comparison of the MDi model (crosses) and the model of \cite{Seiz2023a} (circles) when employing the rectangular grain geometry.
 While the MDi model converges at around 8 particles in the chain, model ADV-$\mu$ does not converge at all.
 The velocity distribution shows that model ADV-$\mu$ tends to produce nonlinear velocity profiles.
 }
 \label{fig:dens-block}
 \end{center}
\end{figure}

The chain ends being different can be somewhat mitigated by employing the rectangular grain geometry from the MD simulations and placing pores on the grain boundaries.
It will not be fully mitigated, as the absorption rate still depends on the average surface chemical potential, which is different for end grains and for inner grains.
However, as  \cref{fig:dens-block} shows the convergence is sped up with this geometry for the MDi model, but the lack of convergence for ADV-$\mu$ becomes even more obvious.
It should be noted that the MDi model eliminates the pores on the GBs at similar times, whereas model ADV-$\mu$ eliminates them step by step from the outer parts of the chain, with videos showing this being deposited with the Supplementary Material.
This is also the reason for the different velocity magnitudes between the two models:
Up to the point of pore elimination, the size of the pores in the simulations with model MDi is roughly comparable to that of the outermost pore of ADV-$\mu$ at the same time.
Thus the pores for simulations with model MDi are on average smaller than for model ADV-$\mu$ which implies a larger driving force for vacancy absorption.

Based on these results, one can conclude that at about 16 particles in a one-dimensional chain the model MDi becomes representative.
Presuming that this result extends to three-dimensions, one would expect representative simulations to start from about $16^3 = 4096$ particles.
However, it is likely that the end geometry problem makes this a significant overestimation, and thus we proceed to three-dimensional simulations to test this.

\FloatBarrier

\subsection{Large scale three-dimensional simulations}
\label{sec:3d}
\begin{figure}[p]
 \centering
 \includegraphics[width=\textwidth]{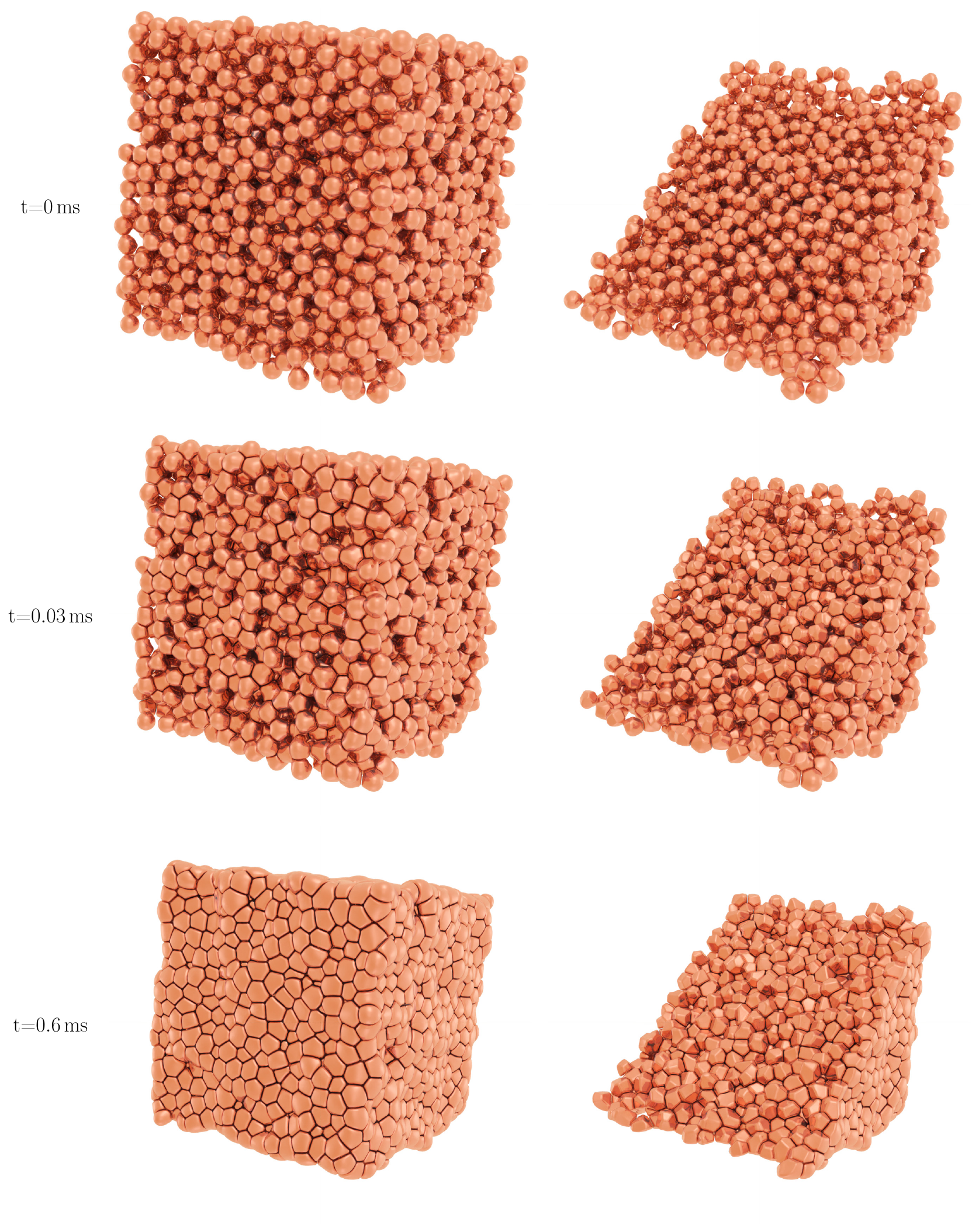}
 \caption{Three-dimensional view of the $400^3$, $R=\SI{12}{nm}$ simulation for various times using Paraview\cite{paraview}.
 The left side shows the entire green body, with the right side showing a (011) fracture surface from the same angle.
 The dark lines delineating regions can be interpreted as grain boundaries and higher order junctions.
 The dark smudges on some grains are due to the dark areas being reflected via ray tracing and thus not actually part of the simulation data.
 Both macroscopic densification and the polyhedralization of the grains are evident.}
 \label{fig:views}
\end{figure}

The model will now be employed to simulate three-dimensional green bodies in order to determine representative volume elements (RVEs) by verifying that the densification is independent of the green body size.
For this, the packings as described in \cite{Seiz2023a} are employed:
A voxel domain of size $N_v^3, N_v \in \{200,400,800\}$ is filled based on a packing generated with the discrete element method.
In order to minimize boundary effects, it is ensured that there are at least 15 voxels between the outermost edge of a particle and the global boundary.
All fields employ zero-flux conditions on the global boundary.
The voxelization happens with a fixed number of voxels $R \in \{8, 12, 16\}$ used to resolve the particle radius, allowing the investigation of particle size effects with different simulations.
Given the employed non-dimensionalization scales and discretization, these correspond to particles of size $R \in \{8, 12, 16\}\si{nm}$.
\Cref{tab:pcounts} lists the number of grains $N_g$ within each combination of $(N_v^3, R)$, \corM{as well as on how many cores the simulations were run and for how long}.
\Cref{fig:views} shows one of the structures at different simulation times; videos of the entire process are deposited with the Supplementary Material.
The left side shows a simple visualization of the entire green body, with the right side showing a fracture surface generated with a (011) plane and removing any grains beyond this plane.
The mesh visualized here is based on a cellwise maximal value of the phase-field vector excepting $\phi_\Vap$.
Contour levels $l > 0.5$ of this field cause an etching-like effect to appear starting from the highest order junctions\footnote{This can also cause grains on the edge of the packing to look only tenously or not connected at all.}.
A contour level of $l=0.6$ is employed which entirely reveals the triple lines in the three-dimensional structure.
In any case, both visualizations show the macroscopic densification of the body, with the fracture surface view also showing that the grains transform from spheres to polyhedra.

\begin{table}
\centering
 \caption{Initial grain counts $N_g$ for the employed packings \corM{ as well as the number of cores $C$ employed and the total runtime $T$.
 The longer runtime of larger particles is due to these being run for longer to achieve comparable densities.}}
\label{tab:pcounts}
 \begin{tabular}{c | c | c | c | c}
  $N_v^3$ & $R$ & $N_g$ & \corM{$C$}  & \corM{$T$}\\\hline
$200^3$ & \SI{16}{ nm} & 97    &  \corM{128 } & \corM{\SI{10.5}{h}}\\
$200^3$ & \SI{12}{ nm} & 262   &  \corM{128 } & \corM{\SI{8.00}{h}}\\
$200^3$ & \SI{8 }{nm } & 1033  &  \corM{128 } & \corM{\SI{4.70}{h}} \\
$400^3$ & \SI{16}{ nm} & 1361  &  \corM{512 } & \corM{\SI{21.7}{h}}  \\
$400^3$ & \SI{12}{ nm} & 3445  &  \corM{512 } & \corM{\SI{16.0}{h}} \\
$400^3$ & \SI{8 }{nm } & 12418 &  \corM{512 } & \corM{\SI{15.6}{h}}  \\
$800^3$ & \SI{16}{ nm} & 14113 &  \corM{8192} & \corM{\SI{15.0}{h}} \\
$800^3$ & \SI{12}{ nm} & 34459 &  \corM{8192} & \corM{\SI{9.20}{h}}  \\
$800^3$ & \SI{8 }{nm } & 120132 & \corM{8192} & \corM{\SI{10.5}{h}}  \\
 \end{tabular}

\end{table}

\begin{figure}[h]
\centering
     \begin{subfigure}[]{0.9\textwidth}
    \centering
        \includegraphics[width=\textwidth]{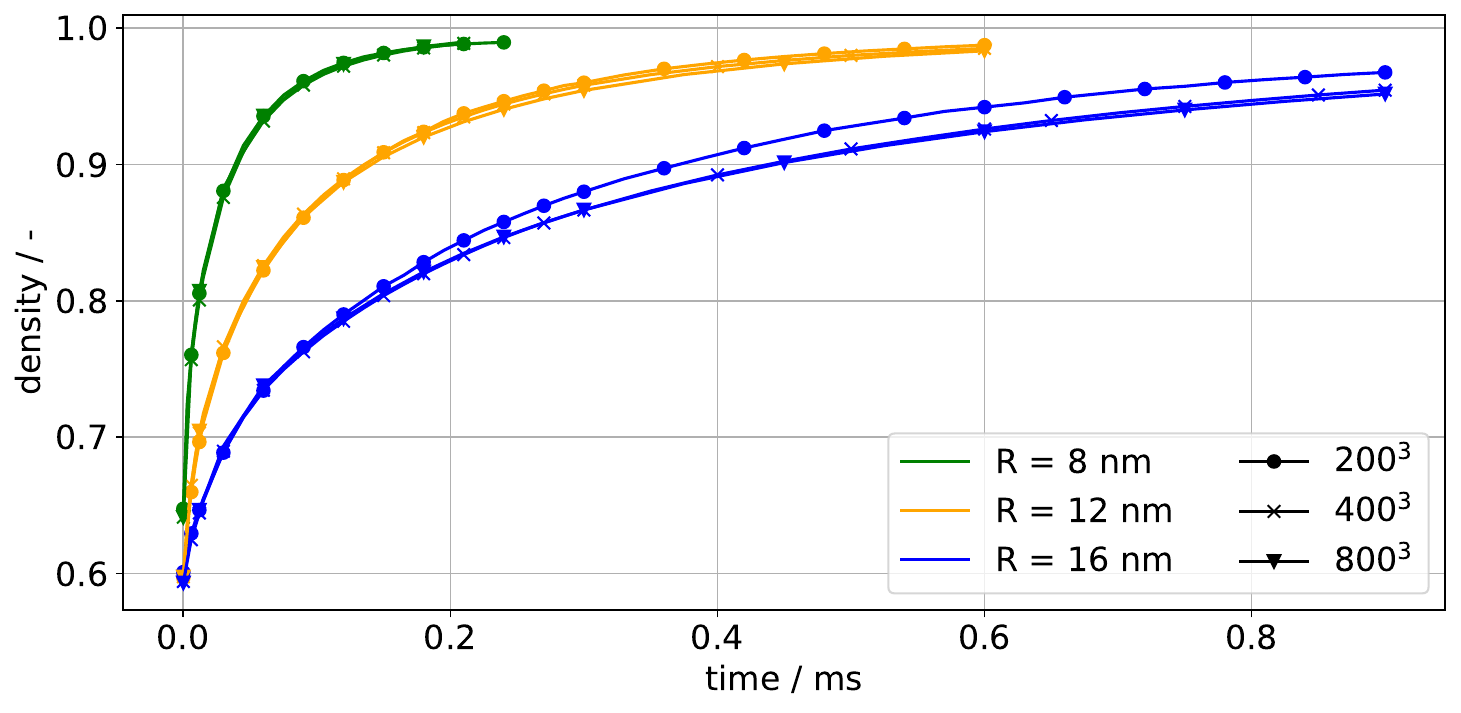}
        \caption{present model MDi}
        \label{fig:mddens}
    \end{subfigure}
~
    \begin{subfigure}[]{0.9\textwidth}
    \centering
        \includegraphics[width=\textwidth]{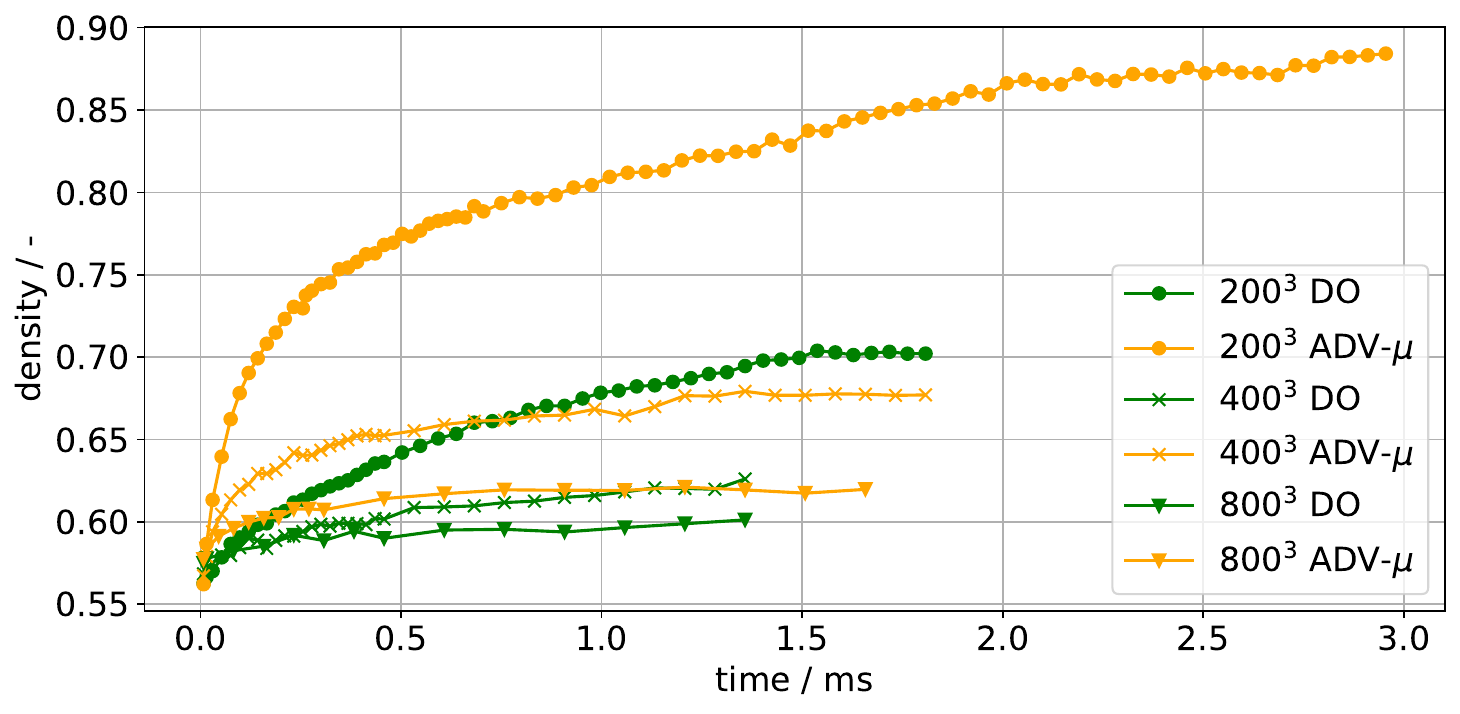}
        \caption{results of \cite{Seiz2023a} ($R = \SI{12}{nm}$)}
        \label{fig:wangdens}
    \end{subfigure}
 \caption{Comparison of densification for the present MDi model and the results of \cite{Seiz2023a} for a diffusion-only (DO) model and a model including advection (ADV-$\mu$) based only on nearest-neighbour interactions.
 The results of \cite{Seiz2023a} clearly have a strong dependence on domain size, which the MDi model eliminates once a RVE is reached.}
 \label{fig:densi}
\end{figure}

The evolution of the density is shown in \cref{fig:densi}, with the present model as well as the results of \cite{Seiz2023a} for a particle size of \SI{12}{nm}.
It can easily be seen that the present model shows highly similar density evolution over all the considered green body sizes.
Only for $R=16$ there is a slight effect of green body size from $200^3$ to $400^3$, with the following simulations being quite similar again.
It is likely that rather than the green body size, a certain minimum number of particles should be contained within a three-dimensional packing, with that threshold lying between 97 particles ($200^3$, $R = \SI{16}{nm}$) and 262 particles ($200^3$, $R = \SI{12}{nm}$).
The particle count likely acts as a proxy variable for whether the geometry is sufficiently homogeneous.
Thus green bodies with a polydispserse grain size distribution will likely need a larger number of particles to be representative, but this will be left to future work.
Furthermore, it is easily seen that densification progresses more slowly with larger grains.

Let us shortly revisit why the present model does not fail to reach a RVE:
The necessary requirement for densification is for the divergence of velocity $\div v$ to be negative between volume elements.
In advection models only employing nearest-neighbour interactions such as \cite{Wang2006} and models based on it, the velocity of particles only depends on their immediate neighbours.
Since \corM{within the green body proper, a grain's immediate neighbours} will be similar, the neighbouring volume elements will be similar, with the exception of those volume elements containing the green body boundary.
In contrast, in the MDi model the velocity of a single particle depends on \emph{all} particles via solving  \cref{eq:contact}.
Thus there is nothing forcing neighbouring volume elements to be similar.
The simplification of using a linear ansatz for the particle displacement of course forces a constant $\div v$ between neighbouring volume elements.
Given that solving the complete system for spatially uncorrelated absorption does result in the particle displacement being a linear function of position, this is quite justified.

\begin{figure}
\includegraphics[width=\textwidth]{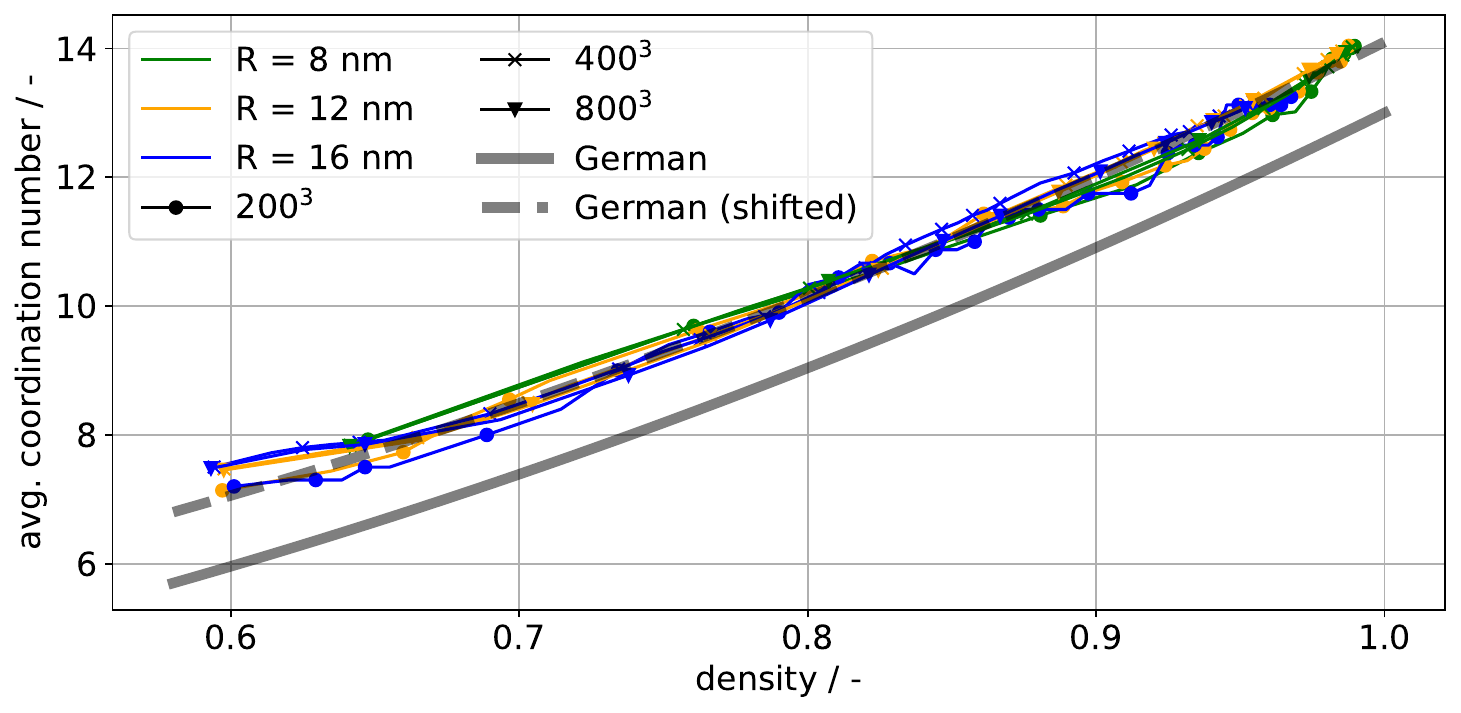}
 \caption{
 The Coordination number over density for all simulations as well as the relation of German \cref{eq:german} are depicted.
 \corM{The coordination number rises monotonically with density, but shows a systematic deviation from German's relation.
 However, the slope of the curve is highly similar, as is shown by also plotting a vertically shifted version of German's relation.
 }
 }
 \label{fig:neighs}
\end{figure}

The present results also allow to qualitatively test whether the geometries assumed in intermediate stage sintering are found.
Coble's classical model \cite{Coble1961} assumes a tetrakaidecahedron, i.e. a solid with 14 faces, for the grain shape.
Thus the number of neighboring grains, also called the coordination number $N_c$, should tend towards 14.
A more quantitative relation is given by German \cite{German2014b}, based on a fit of literature data:
\begin{align}
 N_c = 2 + 11f^2 \label{eq:german}
\end{align}
with the fractional density $f$ which is equivalent to the present usage of density.
The average coordination number of all grains is plotted over density in \cref{fig:neighs}, \corM{showing a monotonic increase of coordination number with density}.
\corM{
After correcting for surface effects, a coordination number of 14 is reached at around $\SI{99}{\percent}$ density.
This includes the effect of many small contacts, which might not be detected easily in experiments.
Hence there is a systematic deviation from German's fit, but the slope is quite comparable.
If the fit is shifted vertically by a value of 1.1, which is roughly the difference in starting coordination number at $\SI{60}{\percent}$ density, then a quite close match is observed.
Conversely, the simulation data could also be filtered to exclude contacts with small surface area, also resulting in a reasonable match for higher densities.
This is explored in the Supplementary Material, as choosing any one minimum surface area is quite arbitrary.
}
In any case, based on both the experimental fit due to German and the present results, the tetrakaidecahedron shape assumption does not hold in intermediate stage sintering.
\corM{It can however hold in the final stage.}

\section{Conclusion}

In the present paper molecular dynamics (MD) is employed to investigate the densification behaviour of a chain of grains.
The grains are observed to move largely as rigid bodies, i.e. the atomic displacement within a single grain is largely uncorrelated to atomic position.
Three rules of motion for this displacement are found:
The displacement is proportional to the number of absorbed vacancies, antiproportional to the grain boundary area and superimposable if multiple vacancy absorption sites are present.
These rules were used to construct an analytical model which agreed well with the MD results.
Following this, a previously published phase-field model (ADV-$\mu$) was extended with this new model for calculating velocities (MDi).
The previously published phase-field model and the new model are then compared in terms of their strain evolution within a linear chain geometry.
For the MDi model, the strain as a function of time is observed to become independent of the number of grains between 8 and 16 grains in the chain, depending on the particulars of the geometry.
Model ADV-$\mu$ did not converge, as previously shown by \cite{Seiz2023a}.
Finally, the model MDi is employed to sinter large-scale three-dimensional structures to determine representative volume elements.
It is found that between 97 and 262 particles are necessary for densification to become independent of the green body size.
Furthermore, the qualitative correct influence of particle size is included in the model, with green bodies consisting of larger particles sintering more slowly.
Finally, reasonable agreement with a model linking the coordination number of grains to the density could be shown.

A future work exploring the applications of the model to pressure-assisted sintering and the investigation of concurrent densification and grain growth is planned.

\section*{Acknowledgements}
The authors thank Andreas Reiter for fruitful discussion on how to efficiently solve \cref{eq:contact} and various other topics.

This work was partially performed on the national supercomputer Hawk at the High Performance Computing Center Stuttgart (HLRS) under the grant number pace3d.
The authors gratefully acknowledge financial support for the modelling of sintering by the Deutsche Forschungsgemeinschaft (DFG) under the grant number NE 822/31-1 (Gottfried-Wilhelm Leibniz prize) and support for the parallelization and code optimization by Karlsruhe Nano Micro Facility (KNMFi) within the programme MSE (P3T1) no. 43.31.01.
%

\section*{Author contributions}
\textbf{Marco Seiz}: Conceptualization, Software, Methodology, Investigation, Data Curation, Validation, Visualization, Resources, Writing - original draft, Writing - review \& editing.
\textbf{Henrik Hierl}: Software, Investigation, Data Curation, Validation, Visualization, Resources, Writing - review \& editing.
\textbf{Britta Nestler}: Funding acquisition, Writing - review \& editing

\section*{Conflicts of interest or competing interests}
The authors declare that there are no conflicts of interest.

\section*{Data and code availability}
The code required to reproduce the present work cannot be shared publicly.
The raw and processed data supporting the findings are available upon reasonable request.

\section*{Supplementary Material}
The Supplementary Material of this paper is available at \url{https://zenodo.org/record/8215289}.
It contains video files of selected simulations as well as a Python implementation for building and solving \cref{eq:contact} for arbitrary connected packings.
The packings employed as initial conditions for the present work are also saved there together with a simple reader for the data.
\corM{Furthermore, the additional analysis of the coordination number and surface area distribution is deposited there as well.}

\section*{Ethics approval}
Not applicable.

%

\bibliography{literatur}

\end{document}